\documentclass[prd,twocolumn,showpacs,amsmath,amssymb,nofootinbib,preprintnumbers]{revtex4}
\usepackage{axodraw}
\usepackage{subfigure}
\usepackage{float}
\usepackage{wrapfig}
\usepackage{array}
\usepackage{graphics}
\usepackage{graphicx}
\usepackage{color}

\newcommand{\fss}[1]{#1\!\!\!/}

\begin{document}

$\text{}$ \preprint{HD-THEP-02-23}

\title{Flow Equations without Mean Field Ambiguity}

\author{Joerg Jaeckel}
\email{Jaeckel@thphys.uni-heidelberg.de}
\author{Christof Wetterich}%
 \email{C.Wetterich@thphys.uni-heidelberg.de}
\affiliation{Institut f\"ur Theoretische Physik, Universit\"at
Heidelberg, Philosophenweg 16, 69120 Heidelberg}

\begin{abstract}
We compare different methods used for non-perturbative
calculations in strongly interacting fermionic systems. Mean field
theory often shows a basic ambiguity related to the possibility to
perform Fierz transformations. The results may then depend strongly on an
unphysical parameter which reflects the choice of the mean field, thus limiting
the reliability. This ambiguity is absent for Schwinger-Dyson equations
or fermionic renormalization group equations. Also
renormalization group equations in a partially bosonized setting
can overcome the Fierz ambiguity if the truncation is chosen
appropriately. This is reassuring since the partially bosonized
renormalization group approach constitutes a very promising basis for the explicit treatment of
condensates and spontaneous symmetry breaking even for situations where the bosonic
correlation length is large.
\end{abstract}
\maketitle
\section{Introduction}
Mean field theory is a widely used method in statistical physics and quantum field
theory, in particular for ground states characterized by condensates and spontaneous symmetry breaking.
For example, mean field solutions of the Nambu-Jona-Lasinio (NJL) model
\cite{Nambu:1961tp} or extensions of it
are one of the main theoretical tools in nuclear physics.
The recent discussion of color superconductivity
at high but realistic baryon density is mainly based on this method
\cite{Alford:1997zt,Bailin:bm,Berges:1998rc,Alford:1998mk,Schafer:1998ef,Oertel:2002pj}.
One out of many examples from statistical physics is a mean field description \cite{Baier:2000yc} of antiferromagnetic
and superconducting condensates in the Hubbard model \cite{Gutzwiller,Kanamori,Hubbardmodell}.
Quite generally, mean field theory (MFT) seems to be well
suited for systems with multifermion interactions and bosonic condensates.

It is well known that MFT has a basic ambiguity which is
connected with the possibility to perform Fierz transformations
(FT) for the underlying local multifermion interaction. For example, in the Hubbard model this ambiguity has
a sizable influence on the results \cite{Baier:2000yc}. The origin of this ambiguity becomes apparent already in the simplest
NJL-type model (for only one fermion species) with a chirally invariant pointlike four fermion interaction:
\begin{eqnarray}
\label{equ::faction}
\textrm{S}_{\textrm{F}}=\int d^{4}x
&\bigg\{&\bar{\psi}i\fss{\partial}\psi
+\frac{1}{2}\lambda_{\sigma}[(\bar{\psi}\psi)^{2}-(\bar{\psi}\gamma^{5}\psi)^2]
\\\nonumber
&-&\frac{1}{2}\lambda_{V}[(\bar{\psi}\gamma^{\mu}\psi)^2]
-\frac{1}{2}\lambda_{A}[(\bar{\psi}\gamma^{\mu}\gamma^{5}\psi)^{2}]\bigg\}.
\end{eqnarray}
In this paper we concentrate on this model which is regularized by a sharp momentum cutoff $q^{2}<\Lambda^{2}$.

Due to the Fierz identity
\begin{equation}
\label{equ::fierz}
\left [(\bar{\psi}\gamma^{\mu}\psi)^{2}-(\bar{\psi}
\gamma^{\mu}\gamma^{5}\psi)^{2}\right]
+2\left[(\bar{\psi}\psi)^{2}-(\bar{\psi}\gamma^{5}\psi)^{2}\right ]=0
\end{equation}
only two of the quartic couplings are independent and we write
\begin{equation}
\label{equ::invariant}
\lambda_{\sigma}=\bar{\lambda}_{\sigma}+2\gamma\bar{\lambda}_{V},
\quad \lambda_{V}=(1-\gamma)\bar{\lambda}_{V},\quad \lambda_{A}=\gamma\bar{\lambda}_{V}.
\end{equation}
The parameter $\gamma$ is redundant since it multiplies just the vanishing expression \eqref{equ::fierz}.
No physical quantity can depend on $\gamma$ in a full computation of the functional integral for the
partition function and expectation values of field operators. The model is completely characterized by
the two ''physical'' couplings $\bar{\lambda}_{\sigma}$ and $\bar{\lambda}_{V}$.

We will see in sect. \ref{sec::mean} that
the MFT results can strongly depend on $\gamma$, limiting their quantitative
reliability. For example, the value of the critical coupling $\bar{\lambda}^{\textrm{crit}}_{\sigma}$
for the onset of a nonvanishing condensate
$\sigma\sim\langle\bar{\psi}\left(\frac{1-\gamma^{5}}{2}\right)\psi\rangle$ depends on $\gamma$
for fixed $\bar{\lambda}_{V}\neq 0$ \mbox{(cf. the} values
in the first row of Tabs. \ref{tab::crit}, \ref{tab::crit2}). This ''Fierz ambiguity'' is a strong
effect unless $|\bar{\lambda}_{V}|$ is much smaller than $|\bar{\lambda}_{\sigma}|$.
Similarly, for given $\bar{\lambda}_{V}$, $\bar{\lambda}_{\sigma}$ the mean
field value of the condensate $\sigma$ in the phase with spontaneous symmetry breaking
depends on $\gamma$.

The origin of the mean field ambiguity is easy to understand: once the pointlike interaction
has been ''distributed'' on the channels $(S-P)$, $(V)$, $(A)$ (with respective couplings
$\lambda_{\sigma}$, $\lambda_{V}$, $\lambda_{A}$) the interactions in the channels
$(V)$ and $(A)$ do not influence\footnote{We refer here to MFT as used in most
computations and implemented on a more formal level by the Hubbard-Stratonovich transformation
or partial bosonization. Sometimes the wording ''mean field'' is also used for a Schwinger-Dyson
approach for which no ambiguity is present. We discuss this in sect. \ref{sec::schwinger}. We note
that the MF-ambiguity gets enhanced once we include, in addition, di-fermion channels
$\sim \psi\psi$ and $\bar{\psi}\bar{\psi}$.} the computation in the channel $(S-P)$
anymore -- at least as long as the mean fields $\sim\bar{\psi}\gamma^{\mu}\psi$ and
$\bar{\psi}\gamma^{\mu}\gamma^{5}\psi$ do not yet get an expectation value.
This distribution depends, however, on the parameter $\gamma$. It may sometimes
be possible to develop from other considerations an educated guess what should be the
distribution of the interaction on the various channels, thus limiting the range of
''acceptable'' $\gamma$. Still, the spread of the results over the acceptable range
of $\gamma$ should be considered as a lower bound for the systematic uncertainty of
a MFT computation. Depending on the other uncertainties the relative importance of this
''Fierz-ambiguity'' may be more or less important. For the example of Tab. \ref{tab::crit} the
Fierz ambiguity of MFT seems to be of the same size as the spread in the results
between the different methods beyond MFT. On the other hand, for the values used in Tab.
\ref{tab::crit2} the ambiguity for the range
$0\leq \gamma\leq 1$ is so large that no quantitative statement is possible for MFT
unless $\gamma$ can be restricted to a much smaller range.

A correct ''guess'' of $\gamma$ is often difficult\footnote{If the effective four fermion
interaction is known beyond the pointlike limit the distribution on the various channels
is much more restricted. It should be approximated by a sum of terms where $\lambda_{V}$
only depends on the total momentum of the bilinear $\bar{\psi}\gamma^{\mu}\psi$ and similar
for the other channels.}. One may therefore prefer to reduce the dependence on the unphysical
parameter $\gamma$ by computing in a more elaborate approximation. This situation is
very similar to perturbative computations in quantum field theory: the results in a
given order depend on the renormalization scheme and the renormalization scale $\mu$.
Typically, the scheme dependence ($\mu$-dependence) gets reduced in higher orders.
Since the physical results cannot depend on $\mu$ the residual $\mu$-dependence is
often used as a guess for the remaining error -- in fact it constitutes a lower
bound for the systematic uncertainty in a given order of the perturbative expansion.

In this paper we want to investigate methods where MFT appears as some type of first step in a more
systematic expansion. We find indeed that for these methods the
dependence on $\gamma$ is reduced as compared to MFT. For
some approximations a residual $\gamma$-dependence remains which should again
be considered as a lower bound on the systematic uncertainties at this
level\footnote{Other uncertainties not related to the Fierz ambiguity may still be larger.
This is obviously the case for methods without a Fierz ambiguity.}.
We will discuss methods based on the
exact renormalization group equation for the effective average action \cite{Wetterich:1993be}
or on Schwinger-Dyson equations \cite{Dyson:1949ha,Schwinger:1951ex}. To improve the approximation
we include additional diagrams similar to those needed at order $\frac{1}{N}$ in the $\frac{1}{N}$-expansion
\cite{Manohar:1998xv,'tHooft:1974hx,'tHooft:1973jz,Gat:xi,Hands:1992be}.

As a guidance, we first study in sect. \ref{sec::perturbation} perturbation theory
in the one-loop approximation. While
the results are independent of $\gamma$, the
validity of the perturbative calculation is limited to small coupling. In particular, the interesting
phenomenon of spontaneous symmetry breaking (SSB) cannot be seen.

This shortcoming is improved substantially by the use of a non-perturbative flow equation
for the scale-dependence of the effective average action $\Gamma_{k}$ \cite{Wetterich:1993be}.
In this approach an infrared cutoff with scale $k$ is introduced such that only fluctuations with
momenta $q^{2}\gtrsim k^{2}$ contribute effectively. The effective couplings therefore depend on
the scale and their \mbox{$k$-dependence} is governed by a renormalization group (RG) equation. For
example, in a simple truncation $\Gamma_{k}$ takes the same form as the action \eqref{equ::faction},
but the couplings become now scale dependent, i.e. $\lambda_{\sigma}\rightarrow\lambda_{\sigma,k}$
etc.. In the limit $k=0$ the infrared cutoff is absent and all fluctuations are included. For $k=0$
the effective average action $\Gamma_{k=0}$ is the generating functional of the 1PI-Greens functions
of the full theory.

The scale dependence of $\Gamma_{k}$ is described by an exact functional differential
\mbox{equation \cite{Wetterich:1993be}.}
Non-perturbative approximations to this exact equation involve a truncation
of the general form of $\Gamma_{k}$. The non-perturbative flow equations for the quartic couplings
$\lambda_{\sigma,k}$, $\lambda_{V,k}$ and $\lambda_{A,k}$ are solved
in sect. \ref{sec::fermion}.
Now the onset of SSB is signaled by a divergence of the quartic couplings $\lambda$.
The fermionic flow equation yields results for the critical couplings which are independent of $\gamma$.
Unfortunately, in contrast to MFT, one cannot easily extend the investigation to the phase with SSB and
compute, for example, the value of the order parameter or the fermion mass gap. This would require
the inclusion of multifermion interactions in the flow (e.g. eight fermion interactions), leading to
high algebraic complexity.

Partial bosonization seems to be the ideal remedy to this
difficulty
\cite{stratonovich,Hubbard:1959ub,Klevansky:1992qe,Ellwanger:1994wy,Alkofer:1996ph,Berges:1999eu}.
In \mbox{sect. \ref{sec::bosonization}} the model
\eqref{equ::faction} is rewritten as an equivalent Yukawa type
model with scalars $\phi$, vectors $V^{\mu}$ and axial vectors
$A^{\mu}$ representing the corresponding fermion bilinears.
Spontaneous symmetry breaking can now be dealt with by computing
the effective potential for $\phi$ and looking for a minimum at
$\phi\neq 0$. For example, a term $\sim \phi^{4}$ stands for an
eight fermion interaction. Unfortunately, partial bosonization
brings back the ''Fierz ambiguity'' of MFT. In fact, an approximation which only
includes the fermionic fluctuations and omits the bosonic
fluctuations is precisely equivalent to MFT. Nevertheless, MFT can now be
considered as a starting point of a more systematic procedure
which also includes the bosonic fluctuations. Here again, a test
for the validity of any approximation to the bosonic fluctuations
should see that physical quantities become independent of
$\gamma$.

In sects. \ref{sec::bosoflow}, \ref{sec::redef} we study the flow
equations for the Yukawa model with action
\begin{eqnarray}
\label{equ::baction}
\nonumber
\textrm{S}_{\textrm{B}}=\int d^{4}x&\bigg\{&i\bar{\psi}\fss{\partial}\psi
+\mu^{2}_{\sigma}\phi^{\star}\phi
+\frac{\mu^{2}_{V}}{2}V_{\mu}V^{\mu}+\frac{\mu^{2}_{A}}{2}A_{\mu}A^{\mu}
\\\nonumber
&+&h_{\sigma}\left[ \bar{\psi}\left(\frac{1+\gamma^{5}}{2}\right)\phi\psi
-\bar{\psi}\left(\frac{1-\gamma^{5}}{2}\right)\phi^{\star}\psi\right]
\\
&-&h_{V}\bar{\psi}\gamma_{\mu}V^{\mu}\psi
-h_{A}\bar{\psi}\gamma_{\mu}\gamma^{5}A^{\mu}\psi\bigg\}.
\end{eqnarray}
With the identification
\begin{equation}
\label{equ::bosocouplings}
\mu^{2}_{\sigma}=\frac{h^{2}_{\sigma}}{2\lambda_{\sigma}},
\quad\mu^{2}_{V}=\frac{h^{2}_{V}}{\lambda_{V}},
\quad\mu^{2}_{A}=\frac{h^{2}_{A}}{\lambda_{A}}
\end{equation}
this model is equivalent to the NJL-type model \eqref{equ::faction}.
We choose a simple truncation for $\Gamma_{k}$ where only the \mbox{($k$-dependent)} couplings appearing in Eq. \eqref{equ::baction}
are retained.
In \mbox{sect. \ref{sec::bosoflow}} we first neglect the fact that new quartic fermion interactions are generated
by the flow. Including the bosonic fluctuations leads to a running of the Yukawa couplings
$h_{k}$ and reduces substantially
the dependence of the results on $\gamma$ as compared to MFT. Still, the
flow equations are not one-loop exact in this truncation, as also reflected by the residual
\mbox{$\gamma$-dependence} (cf. the third row in Tabs. \ref{tab::crit}, \ref{tab::crit2}).
The reason for
the incompleteness at the one-loop level is the omission of quartic fermion couplings which are generated
by the flow for $k<\Lambda$. In \cite{Gies:2002nw} a systematic method has been developed regarding how
these interactions can be eliminated by $k$-dependent field redefinitions, leading to a modification
of the flow equation. We use this method in sect. \ref{sec::redef}. Indeed, the modified
flow equations in the truncation \eqref{equ::baction} are now one-loop exact.
They are equivalent to the fermionic flow equation of
sect. \ref{sec::fermion}.
At this point we have reached a satisfactory starting point for an extension of the flow equation
beyond the four fermion truncation of sect. \ref{sec::fermion} and beyond the MFT-type truncation
of sect. \ref{sec::redef}. Further extensions of the truncation in the bosonic sector are
now straightforward and will be briefly discussed in sect. \ref{sec::extension}. They are, however,
not the main focus of this paper.

We investigate in sect. \ref{sec::schwinger} bosonization from the viewpoint of the Schwinger-Dyson equations
\cite{Dyson:1949ha,Schwinger:1951ex}.
Those, too, can be applied to both the purely fermionic and the partially bosonized setting.
The formulation using the fermionic model \eqref{equ::faction}
is naturally unambiguous even in the simplest approximation. In the bosonic model \eqref{equ::baction} we investigate
two very simple approximations, the simplest of which corresponds to MFT.
The other one takes into account mass corrections due to diagrams with internal boson lines.
In a very simple approximation for the bosonic propagator and vertices it
corresponds to the gap equation derived for the fermionic model.
The results are therefore independent of $\gamma$. Finally, we present in sect. \ref{sec::conclusions}
an overview over the merits and shortcomings of the different methods.
\section{Critical Couplings from Mean Field Theory} \label{sec::mean}
A mean field calculation treats the fermionic fluctuations in a homogenous background of
fermion bilinears $\tilde{\phi}=\langle\bar{\psi}\left(\frac{1-\gamma^{5}}{2}\right)\psi\rangle$,
$\tilde{\phi}^{\star}=-\langle\bar{\psi}\left(\frac{1+\gamma^{5}}{2}\right)\psi\rangle$,
$\tilde{V}_{\mu}=\langle\bar{\psi}\gamma_{\mu}\psi\rangle$ and
\mbox{$\tilde{A}_{\mu}=\langle\bar{\psi}\gamma_{\mu}\gamma^{5}\psi\rangle$.} It seems straightforward
to replace in the four fermion interaction in Eq. \eqref{equ::faction} one factor by the bosonic mean field,
i.e.
\begin{eqnarray}
\label{equ::meanferm}
\nonumber
(\bar{\psi}\psi)^{2}-(\bar{\psi}\gamma^{5}\psi)^{2}
&\rightarrow&2\tilde{\phi}\bar{\psi}(1+\gamma^{5})\psi-2\tilde{\phi}^{\star}\bar{\psi}(1-\gamma^{5})\psi,
\\\nonumber
(\bar{\psi}\gamma_{\mu}\psi)^{2}&\rightarrow&2\tilde{V}_{\mu}\bar{\psi}\gamma^{\mu}\psi,
\\
(\bar{\psi}\gamma_{\mu}\gamma^{5}\psi)^{2}&\rightarrow&2\tilde{A}_{\mu}\bar{\psi}\gamma^{\mu}\gamma^{5}\psi.
\end{eqnarray}
The partition function becomes then a functional of $\tilde{\phi}$, $\tilde{V}_{\mu}$,
$\tilde{A}_{\mu}$,
\begin{equation}
\label{equ::part}
Z[\tilde{\phi},\tilde{V},\tilde{A}]=\int {\mathcal{D}}\bar{\psi}{\mathcal{D}}\psi
\exp\left(-\textrm{S}[\bar{\psi},\psi,\tilde{\phi},\tilde{V},\tilde{A}]\right),
\end{equation}
where S is given by Eq. \eqref{equ::faction}, with the
replacements \eqref{equ::meanferm}. Self-consistency for the expectation values of
the fermion bilinears requires
\begin{eqnarray}
\label{equ::self}
\tilde{\phi}&=&\frac{1}{2}\langle\bar{\psi}(1-\gamma^{5})\psi\rangle
=\frac{1}{2}\lambda^{-1}_{\sigma}\frac{\partial}{\partial\tilde{\phi}^{\star}}\ln Z,
\\\nonumber
\tilde{V}_{\mu}&=&\langle\bar{\psi}\gamma_{\mu}\psi\rangle
=\lambda^{-1}_{V}\frac{\partial}{\partial\tilde{V}^{\mu}}\ln Z,
\end{eqnarray}
and similar for the other bilinear $\tilde{A}_{\mu}$.
Chiral symmetry breaking by a nonzero $\tilde{\phi}$ requires that the ''field equation''
\eqref{equ::self} has a nontrivial solution. We note that $Z[\tilde{\phi},\tilde{V},\tilde{A}]$
corresponds to a one-loop expression for the fermionic fluctuations in a bosonic background.
With $\Gamma^{(\textrm{F})}_{1}=-\ln Z$ the field equation is equivalent to an extremum
of
\begin{equation}
\label{equ::extremum}
\Gamma^{(\textrm{F})}=
\int d^{4}x\left\{2\lambda_{\sigma}\tilde{\phi}^{\star}\tilde{\phi}
+\frac{1}{2}\lambda_{V}\tilde{V}_{\mu}\tilde{V}^{\mu}
+\frac{1}{2}\lambda_{A}\tilde{A}_{\mu}\tilde{A}^{\mu}\right\}
+\Gamma^{(\textrm{F})}_{1}.
\end{equation}
A discussion of spontaneous symmetry breaking in MFT amounts therefore to a calculation
of the minima of $\Gamma^{(\textrm{F})}$.

We note that this calculation can be done equivalently in the Yukawa
theory \eqref{equ::baction}, \eqref{equ::bosocouplings}.
The mapping of the bosonic fields reads $\phi=\left(h_{\sigma}/\mu^{2}_{\sigma}\right)\tilde{\phi}$,
$V_{\mu}=\left(h_{V}/\mu^{2}_{V}\right)\tilde{V}_{\mu}$, \mbox{$A_{\mu}=\left(h_{A}/\mu^{2}_{A}\right)\tilde{A}_{\mu}$.}
Keeping the bosonic fields fixed and performing the remaining Gaussian fermionic functional
integral yield precisely Eq. \eqref{equ::extremum}. Mean field theory therefore corresponds
precisely to an evaluation of the effective action in the partially bosonized Yukawa model
in a limit where the bosonic fluctuations are neglected.

We want to compute here the critical couplings (more precisely, the critical line in the plane of
couplings $\bar{\lambda}_{\sigma}$, $\bar{\lambda}_{V}$) for which a nonzero expectation value
$\phi\neq 0$ indicates the onset of spontaneous symmetry breaking. For this purpose
we calculate the mass term $\sim \phi^{\star}\phi$ in $\Gamma^{(\textrm{F})}$
and look when it turns negative. This defines the critical couplings. We assume here a situation
where the expectation values of other bosonic fields such as $V_{\mu}$ or $A_{\mu}$
vanish in the relevant range of couplings. It is then sufficient to evaluate
$\Gamma^{(\textrm{F})}$ for $V_{\mu}=A_{\mu}=0$.

In a diagrammatic language gaussian integration over the fermionic variables corresponds to evaluating
the diagram of Fig. \ref{fig::mass}. We define our model with a fixed ultraviolet momentum cutoff
$q^{2}<\Lambda^{2}$, such that the MFT-result becomes ($v_{4}=1/(32\pi^{2})$, $x=q^2$):
\begin{figure}[t]
\begin{center}
\includegraphics{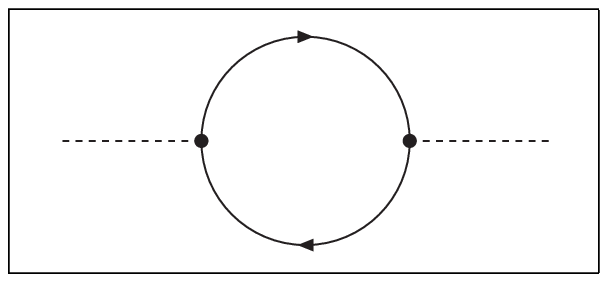}
\end{center}
\caption{Bosonic mass correction due to fermion fluctuations. Fermionic lines are solid with an arrow, bosonic
or ''mean field lines'' $\sim\langle\bar{\psi}\left(\frac{1-\gamma^{5}}{2}\right)\psi\rangle$ are dashed.}
\label{fig::mass}
\end{figure}
\begin{equation}
\label{equ::potential}
\Gamma^{(\textrm{F})}_{1}=-4v_{4}\int^{\Lambda^{2}}_{0}dx\,x\ln(x+h^{2}_{\sigma}\phi^{\star}\phi).
\end{equation}
From this one finds the mean field effective action
\begin{eqnarray}
\Gamma^{(\textrm{F})}&=&\Gamma^{(\textrm{F})}_{0}+\Gamma^{(\textrm{F})}_{1}
\\\nonumber
&=&\left(\mu^{2}_{\sigma}-4v_{4}h^{2}_{\sigma}\Lambda^{2}\right)\phi^{\star}\phi
+\textrm{const}+{\mathcal{O}}\left((\phi^{\star}\phi)^{2}\right),
\end{eqnarray}
where we have expanded in powers of $\phi$.
The mass term turns negative if
\begin{equation}
\label{equ::crit}
\frac{2\mu^{2}_{\sigma}}{h^{2}_{\sigma}\Lambda^{2}}<8v_{4},
\end{equation}
As it should be this result
only depends on the ratio $h^{2}_{\sigma}/\mu^{2}_{\sigma}=2\lambda_{\sigma}$.

We now want to determine the critical line in the plane of invariant couplings $\bar{\lambda}_{\sigma}$,
$\bar{\lambda}_{V}$ from the condition  \eqref{equ::crit}, i.e.
\begin{equation}
\lambda^{\textrm{crit}}_{\sigma}=\frac{1}{8v_{4}\Lambda^{2}}.
\end{equation}
Using the relation \eqref{equ::invariant} we infer a linear dependence
of $\bar{\lambda}^{\textrm{crit}}_{\sigma}$ on $\gamma$ whenever $\bar{\lambda}_{V}\neq 0$
\begin{equation}
\label{equ::MFT}
\bar{\lambda}^{\textrm{crit}}_{\sigma}=\frac{1}{8v_{4}\Lambda^{2}}
-2\gamma\bar{\lambda}_{V}.
\end{equation}
(For numerical values see Tabs. \ref{tab::crit} and \ref{tab::crit2}).
This dependence is a major shortcoming of MFT. We will refer to it as the ''Fierz ambiguity''. The
Fierz ambiguity does not only affect the critical couplings but also influences the values of masses,
effective couplings etc..

The origin of the Fierz ambiguity can be traced back to the treatment of fluctuations.
A FT of the type \eqref{equ::fierz} changes the effective mean field.
In a symbolic language a FT maps
$(\bar{\psi}_{a}\psi_{a})(\bar{\psi}_{b}\psi_{b})\rightarrow(\bar{\psi}_{a}\psi_{b})(\bar{\psi}_{b}\psi_{a})$
where the brackets denote contraction over spinor indices and matrices $\sim\gamma_{\mu}$
or $\sim\gamma^{5}$ are omitted.
A mean field $\bar{\psi}_{a}\psi_{a}$, appears
after the FT as $\bar{\psi}_{a}\psi_{b}$. From the viewpoint
of the fluctuations one integrates out different fluctuating fields
before and after the FT.
It is therefore no surprise that all results depend on $\gamma$.

\section{Partial Bosonization} \label{sec::bosonization}
The MFT calculation introduces ''mean fields'' composed of fermion - antifermion
(or fermion - fermion) bilinears.
This is motivated by the fact that in many physical systems the fermions are
not the only relevant degrees of freedom at low energies.
Bosonic bound states become important and may condense.
Examples are Cooper pairs in superconductivity or mesons
in QCD.
For a detailed description of the interplay between fermionic and composite bosonic
fluctuations it seems appropriate to treat both on equal footing by introducing
explicit fields for the relevant composite bosons. This will also shed more light
on the status of MFT.

Partial bosonization \cite{Ellwanger:1994wy,Berges:1999eu,stratonovich,Hubbard:1959ub}
is achieved\footnote{The inclusion of fermionic and bosonic source terms as well
as infrared cutoffs is straightforward and omitted here for simplicity.}
by introducing unit factors in the functional integral for the partition
function
\begin{eqnarray}
\label{equ::partition}
Z&=&\int{\mathcal{D}}\bar{\psi}{\mathcal{D}}\psi\exp\left(-\textrm{S}[\psi]\right)
\\\nonumber
&=&\int{\mathcal{D}}\bar{\psi}{\mathcal{D}}\psi{\mathcal{D}}\phi{\mathcal{D}}V^{\mu}{\mathcal{D}}A^{\mu}
{\mathcal{N}}_{\phi}{\mathcal{N}}_{V}{\mathcal{N}}_{A}
\exp\left(-\textrm{S}[\psi]\right)
\end{eqnarray}
with
\begin{alignat}{3}
\label{equ::factors}
{\mathcal{N}}_{\phi}&=&\exp\bigg[-\mu^{2}_{\sigma}
&\left(\phi^{\star}
+\frac{h_{\sigma}}{2\mu^{2}_{\sigma}}\bar{\psi}(1+\gamma^{5})\psi\right)
\\\nonumber
&&&\times\left(\phi-\frac{h_{\sigma}}{2\mu^{2}_{\sigma}}\bar{\psi}(1-\gamma^{5})\psi\right)\bigg],
\\\nonumber
{\mathcal{N}}_{V}&=&\,\exp\bigg[-\frac{\mu^{2}_{V}}{2}
&\left(V^{\mu}-\frac{h_{V}}{\mu^{2}_{V}}\bar{\psi}\gamma^{\mu}\psi\right)
\\\nonumber
&&&\times\left(V_{\mu}-\frac{h_{V}}{\mu^{2}_{V}}\bar{\psi}\gamma_{\mu}\psi\right)\bigg],
\\\nonumber
{\mathcal{N}}_{A}&=&\,\exp\bigg[-\frac{\mu^{2}_{A}}{2}
&\left(A^{\mu}-\frac{h_{A}}{\mu^{2}_{A}}\bar{\psi}\gamma^{\mu}\gamma^{5}\psi\right)
\\\nonumber
&&&\times\left(A_{\mu}-\frac{h_{A}}{\mu^{2}_{A}}\bar{\psi}\gamma_{\mu}\gamma^{5}\psi\right)\bigg].
\end{alignat}
The action in the ''bosonic language'' is composed of the original fermionic action
and the arguments of the exponentials in ${\mathcal{N}}_{\phi}$, ${\mathcal{N}}_{V}$ and
${\mathcal{N}}_{A}$. Using the relation \eqref{equ::bosocouplings} one finds
that the quartic fermion interaction is cancelled. It is now replaced by mass terms
for the bosons and Yukawa couplings between bosons and fermions as given by the expression
\eqref{equ::baction}.

Since the action, Eq. \eqref{equ::baction}, is
quadratic in the fermionic fields the functional integral over the fermionic degrees of freedom is
Gaussian and can be done in one step. As we have seen in the previous section this precisely
leads to the MFT results. More precisely, we understand now that for different choices of $\gamma$
the MFT treatment leaves out different bosonic fluctuations. In this context we note that the
bosonic potential in the bosonized effective action must be bounded from below. This restricts
the possible couplings to
$\lambda_{\sigma},\lambda_{V}, \lambda_{A}>0$.
In the invariant variables this restriction translates to $\bar{\lambda}_{\sigma}, \bar{\lambda}_{V}>0$
and for $\gamma$ it implies $0<\gamma<1$.
\section{Perturbation Theory} \label{sec::perturbation}
In order to cure the unpleasant dependence of the MFT result on $\gamma$ we will include
part of the bosonic fluctuations in sects. \ref{sec::bosoflow} and \ref{sec::redef}. Some
guidance for the level of approximations needed can be gained from
perturbation theory in the fermionic language.
Since the four fermion vertex is uniquely characterized by $\bar{\lambda}_{\sigma}$ and
$\bar{\lambda}_{V}$ the perturbative result must be independent of $\gamma$ at
any given loop order. The lowest order
corrections to the four fermion couplings are obtained by expanding the one-loop
expression\footnote{The
supertrace $STr$ provides an appropriate minus sign in the fermionic sector (see e.g. \cite{zinn-justin1995}).
Moreover, in the full $\textrm{S}^{(2)}$ matrix we have a term from the $\delta^{2}/\delta\bar{\psi}\delta\psi$
derivative ($\textrm{S}^{(2)}_{\bar{F}F}$) and a term from the $\delta^{2}/\delta\psi\delta\bar{\psi}$,
accounting for a factor of 2 in the language with the normal trace. The trace includes momentum integration
and summation over internal indices.}
\begin{equation}
\label{equ::oneloop}
\Delta\Gamma^{\textrm{(1-loop)}}=\frac{1}{2}STr\left[\ln\left(\textrm{S}^{(2)}\right)\right]
=-Tr\left[\ln\left(\textrm{S}^{(2)}_{\bar{F}F}\right)\right]
\end{equation}
up to order $(\bar{\psi}\psi)^{2}$. The corresponding graphs are shown in the second
panel of Fig. \ref{fig::summ}. From
\begin{alignat}{2}
\Delta&\Gamma^{\textrm{(1-loop)}}
=v_{4}\Lambda^{2}\bigg\{&
\\\nonumber
&[4\lambda^{2}_{\sigma}-4\lambda_{\sigma}(\lambda_{A}-2\lambda_{V})]
\left[\left(\bar{\psi}\psi\right)^{2}-\left(\bar{\psi}\gamma^{5}\psi\right)^{2}\right]&
\\\nonumber
&+[-2\lambda_{\sigma}\lambda_{V}+4(\lambda_{A}-\lambda_{V})\lambda_{V}]
\left[\left(\bar{\psi}\gamma^{\mu}\psi\right)^{2}\right]&
\\\nonumber
&+\left[-\lambda^{2}_{\sigma}+2\lambda_{\sigma}\lambda_{A}
+3\lambda^2_{V}-2\lambda_{A}\lambda_{V}-\lambda^{2}_{A}\right]
\left[\left(\bar{\psi}\gamma^{\mu}\gamma^{5}\psi\right)^{2}\right]\bigg\}.&
\end{alignat}
we can read off the corrections $\Delta\lambda_{\sigma}$,
$\Delta\lambda_{V}$ and $\Delta\lambda_{A}$ to the coupling constants.
In order to establish that our result is independent of $\gamma$ we use the freedom of FT
to bring our results into a standard form, such that
$\frac{\Delta\lambda_{A}}{\Delta\lambda_{V}}=\frac{\gamma}{1-\gamma}$. Inserting next the invariant variables
\eqref{equ::invariant} leads to:
\begin{eqnarray}
\label{equ::pert}
\Delta\bar{\lambda}_{\sigma}&=&4v_{4}\Lambda^{2}
(\bar{\lambda}^{2}_{\sigma}+4\bar{\lambda}_{\sigma}\bar{\lambda}_{V}
+3\bar{\lambda}^{2}_{V}),
\\\nonumber
\Delta\bar{\lambda}_{V}&=&2v_{4}\Lambda^{2}(\bar{\lambda}_{\sigma}+\bar{\lambda}_{V})^{2}.
\end{eqnarray}
In contrast to MFT the result does not depend on $\gamma$.

The perturbative result, Eq. \eqref{equ::pert}, always leads to finite corrections to the coupling constants. Remembering
that in the fermionic language the onset of SSB is marked by a divergence of the coupling constants,
it becomes  clear that we will never get SSB in perturbation theory. No critical couplings can be calculated.
This is a severe shortcoming of perturbation theory which cannot be overcome by calculating higher
loop orders. Only an infinite number of loops can give SSB.
In the next section we establish how a renormalization group treatment can overcome this difficulty
without encountering the Fierz ambiguity of MFT. A calculation of the critical coupling becomes
feasible. Nevertheless, even this RG-treatment has its limitations once the couplings diverge.
In particular, it does
not allow us to penetrate the phase with SSB. In sects. \ref{sec::bosoflow} and \ref{sec::redef} this
shortcoming will be cured by a RG-treatment in the partially bosonized language. In particular,
we will see in sect. \ref{sec::redef} which diagrams are needed in order to maintain the independence
of results on $\gamma$ in analogy to perturbation theory.
\section{Renormalization Group for Fermionic Interactions} \label{sec::fermion}
The renormalization group equations for the effective average action
\cite{Wetterich:1993be} are obtained by adding a $k$-dependent
infrared cutoff term $\Delta S_{k}$ to the action. The effective average action
$\Gamma_{k}$ results\footnote{We do not differentiate here the notation between the arguments
of $\Gamma_{k}$ and the fluctuating fields. In terms of the arguments of $\Gamma_{k}$ the cutoff
is substracted in the definition of $\Gamma_{k}$ \cite{Wetterich:1993be}.} from a Legendre transform of $\ln Z_{k}$ \cite{Wetterich:1993be}.
Due to the infrared cutoff, $\Gamma_{k}$ receives only contributions from fluctuations with $q^2\gtrsim k^2$.
We take $\Delta\textrm{S}_{k}$ quadratic in the fermion fields,
\begin{equation}
\label{equ::regulator}
\Delta\textrm{S}_{k}=\int \frac{d^{4}q}{(2\pi)^{4}}\bar{\psi}(q)R_{k}(q)\psi(q)
\end{equation}
with $R_{k}$ vanishing fast for $q^{2}\gg k^{2}$.
For a cutoff function which diverges for $k\rightarrow\infty$
and vanishes for $k\rightarrow 0$ the functional $\Gamma_{k}$ interpolates between the
''classical action'' \mbox{$\Gamma_{\infty}=\textrm{S}_{\textrm{F}}$} and the full
effective action $\Gamma_{0}=\Gamma$ which
includes all quantum fluctuations. The \mbox{$k$-dependence} of $\Gamma_{k}$ is governed by the following exact
equation ($t=\ln(k)$):
\begin{eqnarray}
\label{equ::flowequation}
\partial_{t}\Gamma_{k}[\psi]&=&\frac{1}{2}STr\left\{\tilde{\partial}_{t}\ln(\Gamma^{(2)}_{k}[\psi]+R_{k})\right\},
\\\nonumber
\tilde{\partial}_{t}&=&\partial_{t}R_{k}\frac{\partial}{\partial R_{k}}.
\end{eqnarray}
with $\Gamma^{(2)}_{k}$ the second functional derivative of $\Gamma_{k}$ with respect to the fermion fields.
Despite its suggestive one-loop form this is a functional differential equation which  cannot be solved exactly.

A first approximation neglects the $k$-dependence of $\Gamma_{k}$ on the right hand side (RHS).
The solution is the perturbative result, Eq. \eqref{equ::oneloop}. As we have seen in the previous
\mbox{sect. \ref{sec::perturbation}} this approximation does not lead to SSB. For a better approximation
we restrict $\Gamma_{k}$ to the terms specified in Eq. \eqref{equ::faction} but take
all couplings explicitly $k$-dependent. In the action \eqref{equ::faction} we have only local
interactions. Expressed in momentum space the four fermion interactions have no
momentum dependence. This is often referred to as the local potential approximation (LPA)
\cite{Tetradis:1993ts,Hasenfratz:1986dm, Morris:1994ki}.

Decomposing the fluctuation matrix $\Gamma^{(2)}_{k}$
according to
\begin{equation}
\Gamma^{(2)}_{k}+R_{k}={\mathcal{P}}+{\mathcal{F}}
\end{equation}
into a field independent part ${\mathcal{P}}$ (inverse propagator) and a field dependent part
${\mathcal{F}}$ we can expand the RHS of Eq. \eqref{equ::flowequation} as follows:
\begin{eqnarray}
\partial_{t}\Gamma_{k}
&=&\frac{1}{2}STr\{\tilde{\partial}_{t}\left(\frac{1}{\mathcal{P}}\mathcal{F}\right)\}
-\frac{1}{4}STr\{\tilde{\partial}_{t}\left(\frac{1}{\mathcal{P}}\mathcal{F}\right)^{2}\}
\\\nonumber
&+&\frac{1}{6}STr\{\tilde{\partial}_{t}\left(\frac{1}{\mathcal{P}}\mathcal{F}\right)^{3}\}
-\frac{1}{8}STr\{\tilde{\partial}_{t}\left(\frac{1}{\mathcal{P}}\mathcal{F}\right)^{4}\}+\cdots.
\end{eqnarray}
This amounts to an expansion in powers of fields and we can compare the coefficients of the four
fermion terms with the couplings specified by Eq. \eqref{equ::faction}. We obtain
a set of ordinary differential equations for the couplings:
\begin{eqnarray}
\label{equ::fermionflow}
\nonumber
\partial_{t}\bar{\lambda}_{\sigma,k}&=&-8v_{4}l^{(F),4}_{1}(s)k^{2}
(\bar{\lambda}^{2}_{\sigma,k}+4\bar{\lambda}_{\sigma,k}\bar{\lambda}_{V,k}
+3\bar{\lambda}^{2}_{V,k}),
\\
\partial_{t}\bar{\lambda}_{V,k}
&=&-4v_{4}l^{(F),4}_{1}(s)k^{2}(\bar{\lambda}_{\sigma,k}+\bar{\lambda}_{V,k})^{2},
\end{eqnarray}
in agreement with \cite{Aoki:1997fh} where the same model has been
studied. The threshold functions $l^{(F),4}_{1}$ are defined in
\cite{Berges:2000ew}. For our actual calculation we use a linear
cutoff\footnote{The threshold functions depend on the precise
choice of the cutoff. For the very simple truncation used in this
paper this dependence can actually be absorbed by a suitable
rescaling of $k$, cf. App. \ref{app::cutoff}.} \cite{Litim:2000ci} and adapt the threshold functions to
our setting with fixed momentum cutoff $q^{2}<\Lambda^{2}$ in App.
\ref{app::cutoff}. The dependence on $s=k^{2}/\Lambda^{2}$ becomes
relevant only for $k>\Lambda$ whereas for $k<\Lambda$ one has
constants $l^{(F),4}_{1}=1/2$.

The fermionic flow equations\footnote{As discussed above, the perturbative result, Eq. \eqref{equ::pert}, can be recovered
from Eq. \eqref{equ::fermionflow} if we neglect
the $k$-dependence of the couplings on the RHS and perform the
$t$-integration.} \eqref{equ::fermionflow} do not depend on $\gamma$.
In a diagrammatic language we again have evaluated the diagrams of Fig. \ref{fig::summ} (second panel)
but now with \mbox{$k$-dependent} vertices. In the RG-formulation we only go a tiny step $\Delta k$, and reinsert
the resulting couplings (one-loop diagrams) before we go the next step.
This leads to a resummation of loops.
Since Eq. \eqref{equ::fermionflow} is now nonlinear (quadratic terms on the RHS) the couplings
can and do diverge for a finite $k$ if the initial couplings are large enough. Therefore
we observe the onset of SSB and find a critical coupling.
Since Eq. \eqref{equ::fermionflow} is invariant this critical coupling does not
depend on $\gamma$! Values for the critical coupling obtained by numerically solving
\mbox{Eq. \eqref{equ::fermionflow}} can be found in Tabs. \ref{tab::crit} and \ref{tab::crit2}.

The next step in improving this calculation in the fermionic language would
be to take the momentum dependence of the couplings
into account (e.g. \cite{Meggiolaro:2001kp}) or to include higher orders of the fermionic
fields into the truncation. This seems quite complicated and at first sight we have no physical
guess what is relevant. The renormalization group treatment of the bosonic formulation in
sect. \ref{sec::bosonization} seems much more promising in this respect.

\section{Bosonic flow} \label{sec::bosoflow}
The flow equations in the bosonic language are obtained in complete analogy with the
fermionic formulation. In this paper we restrict the discussion to a ''pointlike''
truncation as given by Eq. \eqref{equ::baction} with $k$-dependent couplings.
We will see that in this approximation we reproduce the result of the last section if
we take care of the fact that new fermionic interactions are generated by the flow and
have to be absorbed by an appropriate \mbox{$k$-dependent} redefinition of the bosonic fields.
The crucial advantage of the bosonic formulation is that it can easily be extended. For
example, the bosonic bound states become dynamical fields if we allow for appropriate
kinetic terms in the truncation, i.e.
\begin{eqnarray}
\label{equ::kinetic}
\nonumber
\Delta\Gamma_{\textrm{kin}}\!\!\!&=&\!\!\!\!\int\! d^{4}x\bigg\{Z_{\phi}\partial_{\mu}\phi^{\star}\partial^{\mu}\phi
+\frac{Z_{V}}{4}V_{\mu\nu}V^{\mu\nu}
+\frac{Z_{A}}{4}A_{\mu\nu}A^{\mu\nu}
\\
&&\quad\quad\quad+\frac{Z_{V}}{2\alpha_{V}}(\partial_{\mu}V^{\mu})^2
+\frac{Z_{A}}{2\alpha_{A}}(\partial_{\mu}A^{\mu})^{2}
\bigg\}
\end{eqnarray}
with
\begin{equation}
V_{\mu\nu}=\partial_{\mu}V_{\nu}-\partial_{\nu}V_{\mu},\quad
A_{\mu\nu}=\partial_{\mu}A_{\nu}-\partial_{\nu}A_{\mu}.
\end{equation}
Also spontaneous symmetry breaking can be explicitly studied if we
replace $\mu^{2}_{\sigma}\phi^{\star}\phi$ by an effective potential $U(\phi^{\star}\phi)$
which may have a minimum for $\phi\neq 0$. This approach has been followed in
previous studies \cite{Schaefer:em,Kodama:1999if,Jungnickel:1995fp,Bergerhoff:1999hr}.
In the present paper we will briefly pursue only the extension \eqref{equ::kinetic}
since our main focus is the issue of the Fierz ambiguity.

\begin{figure}[t]
\begin{center}
\includegraphics{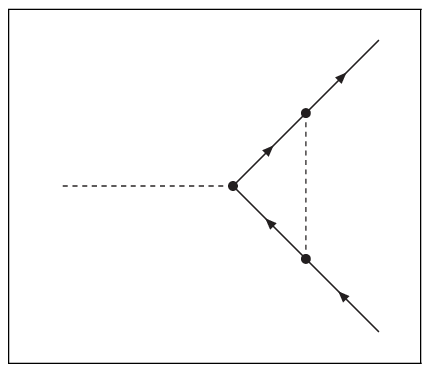}
\end{center}
\caption{Vertex correction diagram in the bosonized model. Solid lines are fermions, dashed lines are bosons.
There exist several diagrams of this type since we have different species of bosons.}
\label{fig::vertex}
\end{figure}

It is instructive to neglect in a first step all bosonic fluctuations by putting
all bosonic entries in the propagator matrix ${\mathcal{P}}^{-1}$ equal to zero. This removes all
diagrams with internal bosonic lines. Among other things this neglects the vertex correction, Fig. \ref{fig::vertex},
and therefore the running of the Yukawa couplings.
Indeed, Fig. \ref{fig::mass} is the only diagram contributing and we recover MFT.
One obtains the flow equations
\begin{eqnarray}
\label{equ::simple}
\nonumber
\partial_{t}\mu^{2}_{\sigma,k}&=&8h^{2}_{\sigma,k}v_{4}k^2l^{(F),4}_{1}(s),
\\\nonumber
\partial_{t}\mu^{2}_{V,k}&=&8h^{2}_{V,k}v_{4}k^{2}l^{(F),4}_{1}(s),
\\\nonumber
\partial_{t}\mu^{2}_{A,k}&=&8h^{2}_{A,k}v_{4}k^{2}l^{(F),4}_{1}(s),
\\
\partial_{t}h_{\sigma,k}&=&0,\quad\partial_{t}h_{V,k}=0,\quad\partial_{t}h_{A,k}=0.
\end{eqnarray}
As long as we do not consider the wave function renormalization \eqref{equ::kinetic} for the bosons,
the flow can be completely described in terms of the dimensionless combinations
\begin{equation}
\label{equ::convenient}
\tilde{\epsilon}_{\sigma,k}=\frac{\mu^{2}_{\sigma,k}}{h^{2}_{\sigma,k}k^{2}},
\quad \tilde{\epsilon}_{V,k}=\frac{\mu^{2}_{V,k}}{h^{2}_{V,k}k^{2}},
\quad \tilde{\epsilon}_{A,k}=\frac{\mu^{2}_{A,k}}{h^{2}_{A,k}k^{2}}.
\end{equation}

Due to the constant Yukawa couplings we can integrate Eq. \eqref{equ::simple}. We find
critical couplings:
\begin{equation}
\label{equ::simplecrit}
\frac{\mu^{2}_{\sigma}}{h^{2}_{\sigma}\Lambda^{2}}\mid_{\textrm{crit}}=4v_{4},
\quad \frac{\mu^{2}_{V}}{h^{2}_{V}\Lambda^{2}}\mid_{\textrm{crit}}=4v_{4},
\quad \frac{\mu^{2}_{A}}{h^{2}_{A}\Lambda^{2}}\mid_{\textrm{crit}}=4v_{4}.
\end{equation}
These are, of course, the results of MFT, Eq. \eqref{equ::crit}.
We note that in Eq. \eqref{equ::simple} the equations for the different species of bosons are completely decoupled.
The mass terms do not turn negative at the same scale for the different species. Indeed it is
possible that the mass of one boson species turns negative while the others do not. Such a behavior is
expected for the full theory, whereas for the fermionic RG of sect. \ref{sec::fermion} all couplings
diverge simultaneously due to their mutual coupling. However, no real conclusion can be taken
from Eq. \eqref{equ::simplecrit} because of the strong dependence on $\gamma$.

Now, let us take also the bosonic fluctuations into account. This includes the vertex correction,
Fig. \ref{fig::vertex}, and the flow of the Yukawa couplings does not vanish anymore.
In the pointlike approximation $(Z_{\phi}=Z_{A}=Z_{V}=0)$ one obtains
\begin{eqnarray}
\nonumber
\partial_{t}h^{2}_{\sigma,k}=
&-&32v_{4}l^{(F),4}_{1}(s)k^{2}h^{2}_{\sigma,k}
\left[\frac{h^{2}_{V,k}}{\mu^{2}_{V,k}}
-\frac{h^{2}_{A,k}}{\mu^{2}_{A,k}}\right],
\\\nonumber
\partial_{t}h^{2}_{V,k}=
&-&4v_{4}l^{(F),4}_{1}(s)k^{2}h^{2}_{V,k}
\\\nonumber
&&\times\left[\frac{h^{2}_{\sigma,k}}{\mu^{2}_{\sigma,k}}
+2\left(\frac{h^{2}_{V,k}}{\mu^{2}_{V,k}}+\frac{h^{2}_{A,k}}{\mu^{2}_{A,k}}\right)\right],
\\\nonumber
\partial_{t}h^{2}_{A,k}=
&-&4v_{4}l^{(F),4}_{1}(s)k^{2}h^{2}_{A,k}
\\
&&\times\left[-\frac{h^{2}_{\sigma,k}}{\mu^{2}_{\sigma,k}}
+2\left(\frac{h^{2}_{V,k}}{\mu^{2}_{V,k}}
+\frac{h^{2}_{A,k}}{\mu^{2}_{A,k}}\right)\right].
\end{eqnarray}
Using the
dimensionless $\tilde{\epsilon}$'s we now find:
\begin{eqnarray}
\label{equ::pubosonic}
\nonumber
\partial_{t}\tilde{\epsilon}_{\sigma,k}&=&-2\tilde{\epsilon}_{\sigma,k}
+8\left[1+4\left(\frac{\tilde{\epsilon}_{\sigma,k}}{\tilde{\epsilon}_{V,k}}
-\frac{\tilde{\epsilon}_{\sigma,k}}{\tilde{\epsilon}_{A,k}}\right)\right]v_{4}l^{(F),4}_{1}(s),
\\\nonumber
\partial_{t}\tilde{\epsilon}_{V,k}&=&-2\tilde{\epsilon}_{V,k}
+8\left[2+\left(\frac{\tilde{\epsilon}_{V,k}}{2\tilde{\epsilon}_{\sigma,k}}
+\frac{\tilde{\epsilon}_{V,k}}{\tilde{\epsilon}_{A,k}}\right)\right]v_{4}l^{(F),4}_{1}(s),
\\\nonumber
\partial_{t}\tilde{\epsilon}_{A,k}&=&-2\tilde{\epsilon}_{A,k}
+8\left[2-\left(\frac{\tilde{\epsilon}_{A,k}}{2\tilde{\epsilon}_{\sigma,k}}
-\frac{\tilde{\epsilon}_{A,k}}{\tilde{\epsilon}_{V,k}}\right)\right]v_{4}l^{(F),4}_{1}(s).
\\
\end{eqnarray}
The onset of spontaneous symmetry breaking is
indicated by a vanishing of $\tilde{\epsilon}$ for at least one species of bosons.
Large $\tilde{\epsilon}$ means that the
corresponding bosonic species becomes very massive and therefore effectively
drops out of the flow.

For initial couplings larger than the critical values (see Tabs. \ref{tab::crit} and \ref{tab::crit2}) both
$\tilde{\epsilon}_{\sigma,k}$ and
$\tilde{\epsilon}_{V,k}$ reach zero for finite $t$. Due to the coupling between the different channels they reach zero
at the same $t$. At this point $\tilde{\epsilon}_{A,k}$ reaches infinity and drops out of the flow. This is quite different
from the flow without the bosonic fluctuations where the flow equations for the different species
were decoupled. The breakdown of all equations at one point resembles\footnote{This is an artefact
of the pointlike approximation.} now the case of the fermionic
model discussed in sect. \ref{sec::fermion}.
The $\gamma$-dependence of the critical couplings  is
reduced considerably, as compared to MFT. This shows that the inclusion of the bosonic fluctuations
is crucial for any quantitatively reliable result. Nevertheless,
the difference between the bosonic and the fermionic flow remains of the order of $10\%$.
\section{Adapted Bosonic Flow} \label{sec::redef}
In our truncation the bosonic propagators are approximated by constants $\mu^{-2}_{k}$. The exchange of bosons
therefore produces effective pointlike four fermion interactions. One therefore
would suspect that this approximation
should contain the same information as the fermionic formulation with pointlike four fermion
interactions. An inspection of the results in Tabs. \ref{tab::crit}, \ref{tab::crit2} shows, however,
that this is not the case for the formulation of the preceding section. In particular,
in contrast to the fermionic language the results of the bosonic flow equations still
depend on the unphysical parameter $\gamma$.

\begin{figure}[t]
\begin{center}
\subfigure[]{
\scalebox{0.96}[0.96]{\includegraphics{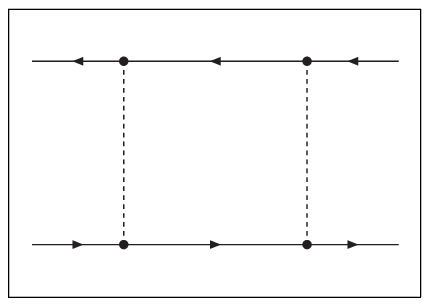}}
}
\subfigure[]{
\scalebox{0.96}[0.96]{\includegraphics{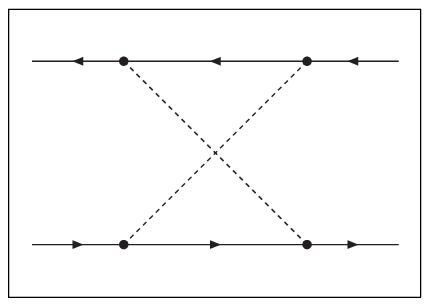}}
}
\end{center}
\caption{Box diagrams for the bosonized model. Again, solid lines are fermions, dashed lines bosons and vertices
are marked with a dot. The diagrams generate new four fermion interactions even for the model \eqref{equ::baction} without
direct four fermion interactions.}
\label{fig::box}
\end{figure}

In fact, even for small couplings $\lambda$ the bosonic flow equations of sect. \ref{sec::bosoflow}
do not reproduce the perturbative result. The reason is that at the one-loop level new quartic
fermion interactions are generated by the box diagrams shown in Fig. \ref{fig::box}.
An easy inspection shows that they contribute in the same order $\lambda^{2}$ as the diagrams
in Figs. \ref{fig::mass} and \ref{fig::vertex}. Even if we start from vanishing
quartic couplings after partial bosonization, such couplings are generated by the flow. The
diagrams in Fig. \ref{fig::box} yield
\begin{eqnarray}
\label{equ::lambdaflow}
\nonumber
\partial_{t}\lambda_{\sigma,k}&=&\beta_{\lambda_{\sigma}}=
-8v_{4}l^{(F),4}_{1}(s)k^{2}\frac{h^{2}_{\sigma,k}}{\mu^{2}_{\sigma,k}}\frac{h^{2}_{A,k}}{\mu^{2}_{A,k}}
+4k^{-2}\tilde{\gamma}(k),
\\\nonumber
\partial_{t}\lambda_{V,k}&=&\beta_{\lambda_{V}}=
24v_{4}l^{(F),4}_{1}(s)k^{2}\frac{h^{2}_{V,k}}{\mu^{2}_{V,k}}\frac{h^{2}_{A,k}}{\mu^{2}_{A,k}}
-2k^{-2}\tilde{\gamma}(k),
\\
\partial_{t}\lambda_{A,k}&=&\beta_{\lambda_{A}}=
-v_{4}l^{(F),4}_{1}(s)k^{2}
\\\nonumber
&&\times\left[
\frac{h^{4}_{\sigma,k}}{\mu^{4}_{\sigma,k}}
-12\frac{h^{4}_{V,k}}{\mu^{4}_{V,k}}
-12\frac{h^{4}_{A,k}}{\mu^{4}_{A,k}}\right]+2k^{-2}\tilde{\gamma}(k).
\end{eqnarray}
Here $\tilde{\gamma}(k)$ is an in principle arbitrary function of scale determining the choice of FT
for the generated four fermion interactions. We will make a special
choice of this function (similar to the
one made in sects. \ref{sec::perturbation} and \ref{sec::fermion}) namely we require
\begin{equation}
\label{equ::fix}
\frac{\tilde{\epsilon}_{V,k}}{\tilde{\epsilon}_{A,k}}=\frac{\gamma}{1-\gamma}\quad \forall \,k,
\end{equation}
with $\tilde{\epsilon}$ given in Eq. \eqref{equ::convenient}. The resulting equation
$\partial_{t}(\tilde{\epsilon}_{V,k}/\tilde{\epsilon}_{A,k})=0$ fixes $\tilde{\gamma}(k)$. An improved
choice of $\tilde{\gamma}(k)$ can be obtained once the momentum dependence of vertices is considered
more carefully \cite{Gies:2002nw}.

An inclusion of the couplings $\lambda_{k}$ into the truncation of the effective average action does not
seem very attractive. Despite the partial bosonization we would still have to deal with the multi-fermion
interactions and the bosonic formulation would be of even higher algebraic complexity
than the fermionic formulation.
A way out of this has been proposed in \cite{Gies:2002nw}. There, it has been shown that it is possible
to reabsorb all four fermion interactions generated during the flow by a redefinition of the bosonic
fields. In the following brief description of this method we use a very symbolic notation.

Introducing an explicit $k$-dependence for the definition of the
bosonic fields in terms of fermion bilinears, the flow equation \mbox{Eq. \eqref{equ::flowequation}}
is modified:
\begin{equation}
\partial_{t}\Gamma_{k}=\partial_{t}\Gamma_{k}\mid_{\phi_{k}}
+\frac{\delta\Gamma_{k}}{\delta\phi_{k}}\partial_{t}\phi_{k}.
\end{equation}
Here $\partial_{t}\Gamma_{k}\mid_{\phi_{k}}=\partial_{t}\Gamma_{k}\mid$
is the standard flow of the effective average action
at fixed fields. Shifting $\phi$ by
\begin{equation}
\partial_{t}\phi_{k}=\left(\bar{\psi}\psi\right)\partial_{t}\omega_{k}
\end{equation}
we find
\begin{eqnarray}
\label{equ::influence}
&&\partial_{t}\mu^{2}=\partial_{t}\mu^{2}\mid,
\\\nonumber
&&\partial_{t}h=\partial_{t}h\mid+\mu^{2}\partial_{t}\omega_{k},
\quad\partial_{t}\lambda=\partial_{t}\lambda\mid-h\partial_{t}\omega_{k}
\end{eqnarray}
and we can choose $\omega_{k}$ to establish:
\begin{equation}
\partial_{t}\lambda=0.
\end{equation}
Instead of including running four fermion couplings explicitly we therefore only have to use
adapted flow equations for the couplings contained in Eq. \eqref{equ::baction}.

Let us now apply this method explicitly to our model. Shifting
\begin{eqnarray}
\label{equ::shift}
&&\partial_{t}\phi=-\bar{\psi}\left(\frac{1-\gamma^{5}}{2}\right)\psi\partial_{t}\omega_{\sigma,k},
\\\nonumber
&&\partial_{t}\phi^{\star}=\bar{\psi}\left(\frac{1+\gamma^{5}}{2}\right)\psi\partial_{t}\omega_{\sigma,k},
\\\nonumber
&&\partial_{t}V^{\mu}=-\bar{\psi}\gamma^{\mu}\psi\partial_{t}\omega_{V,k},
\quad
\partial_{t}A^{\mu}=-\bar{\psi}\gamma^{\mu}\gamma^{5}\psi\partial_{t}\omega_{A,k}
\end{eqnarray}
we have
\begin{eqnarray}
\label{equ::lambda}
\partial_{t}\lambda_{\sigma,k}&=&\partial_{t}\lambda_{\sigma,k}|-h_{\sigma,k}\partial_{t}\omega_{\sigma,k},
\\\nonumber
\partial_{t}\lambda_{V,k}&=&\partial_{t}\lambda_{V,k}|-2h_{V,k}\partial_{t}\omega_{V,k},
\\\nonumber
\partial_{t}\lambda_{A,k}&=&\partial_{t}\lambda_{A,k}|-2h_{A,k}\partial_{t}\omega_{A,k}.
\end{eqnarray}
Requiring $\partial_{t}\lambda=0$ for all $\lambda$'s we can determine the functions $\omega$:
\begin{eqnarray}
\label{equ::omegas}
\partial_{t}\omega_{\sigma,k}=\frac{\beta_{\lambda_{\sigma}}}{h_{\sigma,k}},
\quad
\partial_{t}\omega_{V,k}=\frac{\beta_{\lambda_{V}}}{2h_{V,k}},
\quad
\partial_{t}\omega_{A,k}=\frac{\beta_{\lambda_{A}}}{2h_{A,k}}
\end{eqnarray}
with the $\beta$-functions given in Eq. \eqref{equ::lambdaflow}. This yields the adapted flow
equations for the Yukawa couplings
\begin{eqnarray}
\label{equ::effect}
\partial_{t}h_{\sigma,k}&=&\partial_{t}h_{\sigma,k}|+\mu^{2}_{\sigma,k}\partial_{t}\omega_{\sigma,k},
\\\nonumber
\partial_{t}h_{V,k}&=&\partial_{t}h_{V,k}|+\mu^{2}_{V,k}\partial_{t}\omega_{V,k},
\\\nonumber
\partial_{t}h_{A,k}&=&\partial_{t}h_{A,k}|+\mu^{2}_{A,k}\partial_{t}\omega_{A,k}.
\end{eqnarray}
Combining
Eqs. \eqref{equ::pubosonic}, \eqref{equ::lambdaflow}, \eqref{equ::fix}, \eqref{equ::omegas}, \eqref{equ::effect}
determines $\tilde{\gamma}(k)$
\begin{eqnarray}
\tilde{\gamma}(k)&=&2v_{4}l^{(F),4}_{1}(s)
\\\nonumber
&\times&\left[-\frac{3}{\tilde{\epsilon}^{2}_{V,k}}
+\frac{1}{\tilde{\epsilon}_{V,k}\tilde{\epsilon}_{A,k}}
+\frac{(4\tilde{\epsilon}_{\sigma,k}-\tilde{\epsilon}_{A,k})^{2}}
{4\tilde{\epsilon}^{2}_{\sigma,k}\tilde{\epsilon}_{A,k}(\tilde{\epsilon}_{V,k}+\tilde{\epsilon}_{A,k})}
\right].
\end{eqnarray}

Having fixed the ratio between $\tilde{\epsilon}_{V,k}$ and $\tilde{\epsilon}_{A,k}$
we need only two equations to describe the flow. We will use the ones for $\tilde{\epsilon}_{\sigma,k}$
and $\bar{\epsilon}_{V,k}=(1-\gamma)\tilde{\epsilon}_{V,k}$
\begin{eqnarray}
\label{equ::invariantbos}
\nonumber
\partial_{t}\tilde{\epsilon}_{\sigma,k}=-2\tilde{\epsilon}_{\sigma,k}
&+&4[(1+\gamma)-4(-2+\gamma+2\gamma^2)
\frac{\tilde{\epsilon}_{\sigma,k}}{\bar{\epsilon}_{V,k}}
\\\nonumber
&+&4(3-7\gamma+4\gamma^3)
\frac{\tilde{\epsilon}^{2}_{\sigma,k}}{\bar{\epsilon}^{2}_{V,k}}]l^{(F),4}_{1}(s)v_{4},
\\\nonumber
\partial_{t}\bar{\epsilon}_{V,k}=-2\bar{\epsilon}_{V,k}
&+&4[\frac{\bar{\epsilon}_{V,k}}{2\tilde{\epsilon}_{\sigma,k}}-(2\gamma-1)]^{2}l^{(F),4}_{1}(s)v_{4}.
\\
\end{eqnarray}
These equations are completely equivalent to the fermionic flow Eq. \eqref{equ::fermionflow}.
In order to see this we remember that the simple truncation
of the form \eqref{equ::baction} is at most quadratic in the bosonic fields. We can therefore easily solve
the bosonic field equations as a functional of the fermion fields. Reinserting
the solution into the effective average action we obtain the form \eqref{equ::faction} with
the $k$-dependent quartic couplings
\begin{equation}
\bar{\lambda}_{\sigma,k}=\frac{1}{2k^{2}\tilde{\epsilon}_{\sigma,k}}-2\gamma\frac{1}{k^{2}\bar{\epsilon}_{V,k}},
\quad \bar{\lambda}_{V,k}=\frac{1}{k^{2}\bar{\epsilon}_{V,k}}.
\end{equation}
Inserting this into Eq. \eqref{equ::invariantbos} we find Eq. \eqref{equ::fermionflow}, establishing
both the exact equivalence to the fermionic model and the $\gamma$-independence of physical quantities.

\begin{figure*}[t]
\includegraphics{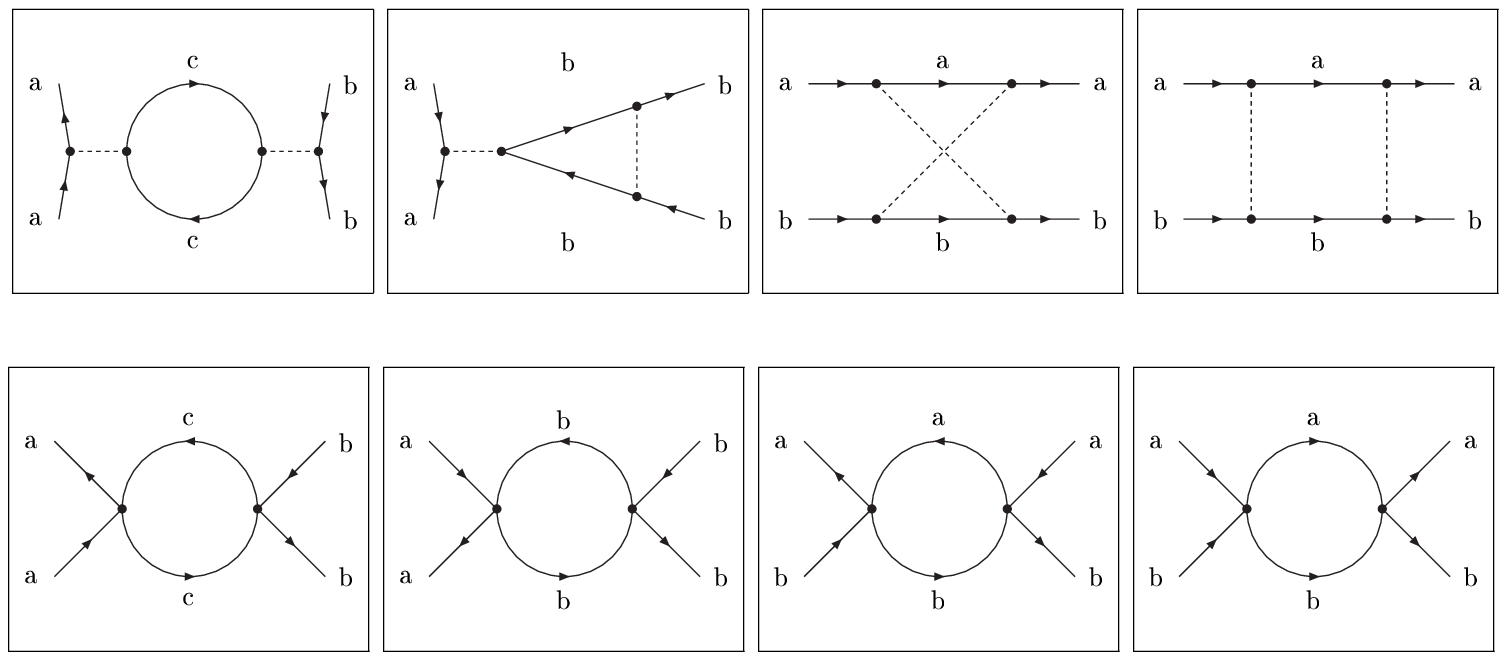}
\caption{Summary of all diagrams encountered in the previous sections.
There is a one to one correspondence between the diagrams of the bosonized model (first row)
and the purely fermionic model
(second row). Solid lines with an
arrow denote fermionic lines. The letters in the diagrams are given to visualize the
ways in which the fermionic operators are contracted, e.g. the first diagram in the second row results from a
term
$[(\bar{\psi}_{a}\psi_{a})
(\bar{\psi}_{c}\psi_{c})][(\bar{\psi}_{c}\psi_{c})(\bar{\psi}_{b}\psi_{b})]$.
Shrinking bosonic lines (dashed) to points maps the diagrams in the first row to the second row.
In the approximations of sect. \ref{sec::bosoflow} only the first or the first two diagrams are taken into account.}
\label{fig::summ}
\end{figure*}

On this level of truncation the equivalence between the fermionic and the adapted bosonic flow can
also be seen on a diagrammatic level.
As long as we do not have a kinetic term for the bosons the internal bosonic lines
shrink to points. On the one-loop level
we find an exact correspondence between the diagrams for the bosonized and the purely fermionic
model summarized in Fig. \ref{fig::summ}. This demonstrates again that one-loop accuracy cannot
be obtained without adaptation of the flow.
\section{The Schwinger-Dyson Approach} \label{sec::schwinger}
Finally, we compare in this section the results of MFT and the RG-treatment with the Schwinger-Dyson (SD)
approach \cite{Schwinger:1951ex,Dyson:1949ha}.
On the exact level the RG and SD approaches are equivalent in the sense that the propagator and higher N-point functions
calculated using the flow \mbox{equation \eqref{equ::flowequation}} are also solutions of the
SD-equations
\cite{Ellwanger:1996wy,Terao:2000ae}. Nevertheless, once truncations are used the results will, in general,
differ. The RG-equations resum diagrams beyond the leading order SD-equation.

\begin{figure*}[t]
\begin{center}
\subfigure[]{
\includegraphics{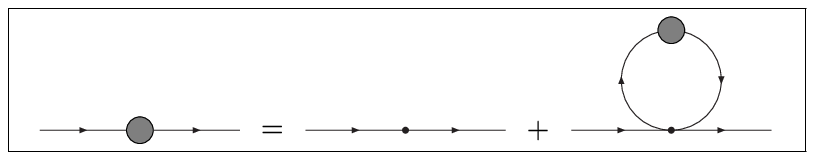}
\label{subfig::schwingerf}
}
\subfigure[]{
\includegraphics{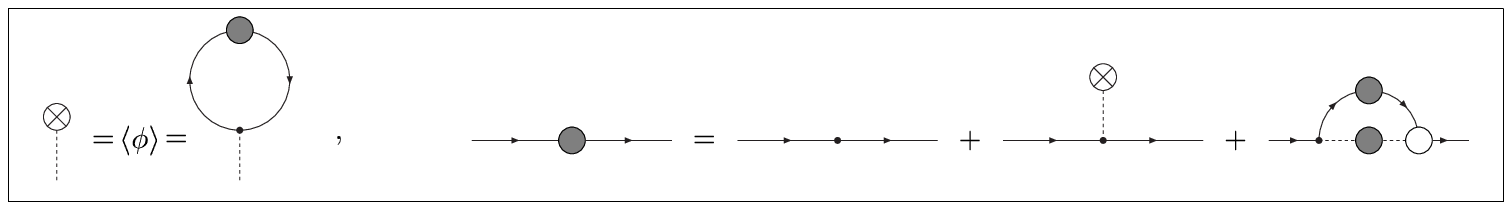}
\label{subfig::schwingerb}
}
\caption{Diagrammatic representation of the lowest order Schwinger-Dyson equations for the fermionic
model (a) (Eq. \eqref{equ::faction}) and the partially bosonized model (b) (Eq. \eqref{equ::baction}).
The shaded circles depict the full propagator, the circle with the cross
is the expectation value of the bosonic field and the empty circle is the full Yukawa vertex.}
\end{center}
\label{fig::schwinger}
\end{figure*}
We start with the purely fermionic model \eqref{equ::faction}.
For this model the Schwinger-Dyson equation, approximated to lowest order,
is depicted in Fig. \ref{subfig::schwingerf}. It is a closed equation since only the bare four
fermion vertex appears. (Only higher order terms involve the full four fermion vertex.)
We write the full fermionic propagator $G_{\textrm{F}}$ as
\begin{equation}
G^{-1}_{\textrm{F}}(p)=G^{-1}_{\textrm{F}0}(p)+\Sigma_{\textrm{F}}(p)
\end{equation}
with the free propagator $G_{\textrm{F}0}$ and self-energy $\Sigma_{\textrm{F}}$. Using this
one obtains a gap equation
for the self-energy which can be solved self-consistently. To simplify
the discussion we make an ansatz for the self-energy:
\begin{equation}
\label{equ::ansatz}
\Sigma_{\textrm{F}}=M_{\textrm{F}}\gamma^{5},
\end{equation}
where the effective fermion mass $M_{\textrm{F}}$ obeys the gap equation
\begin{equation}
\label{equ::complete}
M_{\textrm{F}}=8v_{4}\left[\bar{\lambda}_{\sigma}+\bar{\lambda}_{V}\right]
\int^{\Lambda^2}_{0} dx\,x\frac{M_{\textrm{F}}}{x+M^{2}_{\textrm{F}}}.
\end{equation}
The onset for nontrivial solutions determines the critical couplings:
\begin{equation}
\left[\bar{\lambda}_{\sigma}+\bar{\lambda}_{V}\right]_{\textrm{crit}}=\frac{1}{8v_{4}\Lambda^{2}}.
\end{equation}
This result is shown in Tabs. \ref{tab::crit}, \ref{tab::crit2} and does not depend on $\gamma$,
as expected for a  fermionic calculation. We observe that the MFT-result  for the
$\bar{\lambda}^{\textrm{crit}}_{\sigma}$ coincides with the SD-approach for a particular
choice $\gamma=1/2$. However, in general MFT is not equivalent to the lowest order SD-equation.
This can be seen by computing also the critical coupling for the onset of SSB in
the vector channel. The MFT and SD results do not coincide for the choice \mbox{$\gamma=1/2$.}

Next, we turn to the SD-equations for the bosonized model \eqref{equ::baction} which
are depicted in Fig. \ref{subfig::schwingerb}. We will make here two further approximations
by replacing in the last graph of Fig. \ref{subfig::schwingerb} the full fermion-fermion-boson
vertex by the classical Yukawa coupling and the full bosonic propagator by $\mu^{-2}_{\textrm{B}}$.
We remain with two coupled equations.

In a first step we approximate these equations even further by neglecting
the last diagram in \mbox{Fig. \ref{subfig::schwingerb}} altogether. Then
no fermionic propagator appears on the right hand side of the equation for the fermionic propagator
which only receives a mass correction for $\langle\phi\rangle\neq 0$.
Without loss of generality we take $\phi$ real such that $M_{\textrm{F}}=h_{\sigma}\phi$ and
\begin{equation}
G^{-1}_{\textrm{F}}(q)=-\fss{q}+h_{\sigma}\phi\gamma^{5}.
\end{equation}
Inserting this into the equation for the expectation value $\phi$ we find
\begin{equation}
\label{equ::schMFT}
\phi=\frac{4v_{4}}{\mu^{2}_{\sigma}}\int^{\Lambda^2}_{0} dx\,x
\frac{h^{2}_{\sigma}\phi}{x+h^{2}_{\sigma}\phi^{2}}.
\end{equation}
For the onset of nontrivial solutions we now find the critical value
\begin{equation}
\left[\frac{h^{2}_{\sigma}}{2\mu^{2}_{\sigma}}\right]_{\textrm{crit}}=\frac{1}{8v_{4}\Lambda^{2}}
=\left[\bar{\lambda}_{\sigma}+2\gamma\bar{\lambda}_{V}\right]_{\textrm{crit}}
\end{equation}
which is the (ambiguous) result from MFT given in Eqs. \eqref{equ::crit} and \eqref{equ::MFT}. This is not
surprising since this \emph{exactly is} MFT from the viewpoint of Schwinger-Dyson equations.
Indeed, Eq. \eqref{equ::schMFT} is precisely the field equation which follows by differentiation of the
MFT effective action \eqref{equ::extremum} with respect to $\phi$,
\begin{equation}
\label{equ::MFTpot}
\Gamma^{\textrm{(F)}}=\mu^{2}_{\sigma}\phi^{2}
-4v_{4}\int^{\Lambda^2}_{0}dx\,x\ln(x+h^{2}_{\sigma}\phi^{2}).
\end{equation}

In a next step we improve our approximation to include the full set of diagrams shown in
\mbox{Fig. \ref{subfig::schwingerb}}. Using the same ansatz as before the self-energy
$\Sigma_{\textrm{F}}$ now has two contributions,
\begin{equation}
\label{equ::sum}
\Sigma_{\textrm{F}}=M_{\textrm{F}}\gamma^{5}=h_{\sigma}\phi\gamma^{5}+\Delta m_{\textrm{F}}\gamma^{5}.
\end{equation}
The first one is the contribution due to the expectation value of the bosonic
field whereas $\Delta m_{\textrm{F}}$
is the contribution from the last diagram in Fig. \ref{subfig::schwingerb}, given by
an integral which depends on $M_{\textrm{F}}$. Both in the equation
for $\langle\phi\rangle$ and in the equation for the fermionic propagator only $M_{\textrm{F}}$ appears
on the RHS. Inserting $\langle\phi\rangle$ in the graph, Fig. \ref{subfig::schwingerb}, one finds
a gap equation which determines $M_{\textrm{F}}$:
\begin{equation}
\label{equ::full}
M_{\textrm{F}}=8v_{4}\left[\frac{h^{2}_{\sigma}}{2\mu^{2}_{\sigma}}
+\frac{h^{2}_{V}}{\mu^{2}_{V}}-\frac{h^{2}_{A}}{\mu^{2}_{A}}\right]
\int^{\Lambda^2}_{0}dx\,x\frac{M_{\textrm{F}}}{x+M^{2}_{\textrm{F}}}.
\end{equation}
Once more, this can be expressed in terms of the invariant couplings and again we arrive at
\mbox{Eq. \eqref{equ::complete}}.

Looking more closely at the two contributions to $M_{\textrm{F}}$ we find that alone
neither the contribution $\sim\phi$ (which amounts to MFT as we have discussed above) nor
the ''fermionic contribution'' $\Delta m_{\textrm{F}}$ is invariant under FT's. Only the combination $M_{\textrm{F}}$, which
is the fermion mass and therefore a physical quantity, is invariant. Indeed, changing the
FT amounts to a redefinition of the bosonic fields. In a very similar fashion to what
we have done in sect. \ref{sec::redef} it allows us to choose bosonic fields
such that $\Delta m_{\textrm{F}}=0$. Taking $\gamma=1/2$ gives us such a choice of the bosonic fields. This
explains why MFT gives the same result as the purely fermionic calculation
in this special case.

\section{Beyond the pointlike approximation}\label{sec::extension}
In order to compare the size of the Fierz ambiguity with typical errors from the
truncation we extend in this section our investigation beyond the pointlike approximation.
We add to the general form of the effective action the bosonic propagator terms \eqref{equ::kinetic}
with $\alpha_{V}=\alpha_{A}=1$. In order to simplify the numerical computations we choose in this
section the ERGE-regularization scheme (cf. App. \ref{app::ergescheme}) rather than a sharp
ultraviolet momentum cutoff. As discussed in App. \ref{app::ergescheme} the
definition of the ``classical couplings'' $\bar{\lambda}_{\sigma}$, $\bar{\lambda}_{V}$ differs
from the sharp cutoff regularization.

We have extended the truncation in consecutive steps in order to make the effect of various truncations
directly visible. Our results are shown in Tab. \ref{tab::crit4}. By comparing the first
four rows of Tab. \ref{tab::crit4} with Tab. \ref{tab::crit2} we observe the effect
of the different ultraviolet regularization.
With the ERGE-definition of the ``classical couplings'' one finds a substantially larger value
of the critical coupling $\bar{\lambda}^{\textrm{crit}}_{\sigma}$ than for the sharp
cutoff regularization. This different size is purely a matter of definitions and not
related to any approximation error. Let us now extend the truncation in the partially
bosonized formulation step by step. As discussed above, the lowest order is MFT. As a typical size
for the Fierz ambiguity we quote
$\Delta_{\lambda}=\bar{\lambda}^{\textrm{crit}}_{\sigma}(\gamma=0.75)
-\bar{\lambda}^{\textrm{crit}}_{\sigma}(\gamma=0.25)=-20$. The next step (1) includes the bosonic
fluctuations in the pointlike approximation, without the adaption discussed in sect. \ref{sec::redef}.
At this level the residual Fierz ambiguity is substantially reduced,
$\Delta_{\lambda}=1.71$. Step (2) includes the adaption (box diagrams) in the pointlike
approximation. Due to the equivalence with the pointlike approximation in the purely fermionic
langauge there is no Fierz ambiguity at this level.

As a first step beyond the pointlike approximation we include the
running\footnote{For simplicity we neglect the anomalous dimensions in the threshold
functions.} of the
wave function renormalizations (WFR) $Z_{\phi}$, $Z_{V}$, $Z_{A}$.
(Details of the flow equations will be published elsewhere.) In step (3)
this is done without the adaption by ``rebosonization''. Omitting completely
the rebosonization ((1)+(3)) yields $\Delta_{\lambda}=1.57$ whereas combining Eqs.
(1), (2) and (3) results in $\Delta_{\lambda}=-0.31$. We finally use
rebosonization also for the momentum dependence in the four fermion and Yukawa
couplings generated by the flow (box diagrams and vertex corrections). The results depend somewhat on the detailed
method (to be published elsewhere) and one finds $\Delta_{\lambda}=0.22$. It is impressive
how the truncation reduces the Fierz ambiguity from a value $\Delta_{\lambda}=-20$ for
MFT to a value $|\Delta_{\lambda}|<1$!

On the other hand, a comparison of the last rows in Tab. \ref{tab::crit4} yields a typical
value for the remaining truncation error, $\bar{\lambda}^{\textrm{crit}}_{\sigma}=58\pm 5$.
We emphasize that the quoted error should not be taken as an (more sophisticated)
error estimate in a strict sense. A true error estimate is notoriously difficult in a situation without small
parameters.
One possibility would be further extensions of the truncation. One could also investigate the
influence of different rebosonization procedures or the choice of different cutoff functions $R_{k}(q)$.
Nevertheless, it seems convincing that we have reached a truncation uncertainty that is far less
than the Fierz ambiguity in MFT (i.e. $|\Delta_{\lambda}|=20$). On the other hand, on this
higher level of the truncation the residual Fierz ambiguity is already much less ($|\Delta_{\lambda}|<1$)
than the truncation error. This is precisely what one would like to achieve for a successful non-perturbative
approximation scheme in a partially bosonized setting.

In passing, we note that Schwinger-Dyson equations do a pretty good job in our context.

\section{Summary and Conclusions}\label{sec::conclusions}
We compare different approximation methods for a strongly interacting
fermion system - the NJL model in our case. For this purpose we
have computed the critical couplings for the onset of chiral symmetry breaking using mean
field theory (MFT), fermionic and bosonic renormalization group methods (RG) and the
Schwinger-Dyson equation (SD). This permits a direct comparison between the various methods.
We believe that the general characteristics found here remain valid for other strongly interacting
systems as well. For a sharp momentum cutoff the results for $\bar{\lambda}^{\textrm{crit}}_{\sigma}$ are
summarized in Tabs. \ref{tab::crit}, \ref{tab::crit2} for two fixed values of
$\bar{\lambda}_{V}$. Corresponding results for the ERGE regularization can be found in Tab. \ref{tab::crit4}.
Since the most characteristic features and problems of the different methods are most
clearly seen when the couplings $\bar{\lambda}_{\sigma}$ and $\bar{\lambda}_{V}$ are of similar
size we concentrate the discussion on Tab. \ref{tab::crit2}.

All methods discussed here (except perturbation theory in sect. \ref{sec::perturbation})
correspond to \mbox{non-perturbative}
resummations of perturbative diagrams.
Both MFT and the lowest order SD sum only over fermionic fluctuations in presence of a bosonic background. They include,
in principle, the same type of diagrams, Fig. \ref{fig::mass}. The MFT-result depends
strongly on the choice of the background field. This ''Fierz ambiguity'' is expressed by the
dependence on the unphysical parameter $\gamma$ in the tables. No such ambiguity appears
in the SD approach which therefore seems, at least at first sight, more reliable. We note
that for a particular choice of $\gamma$ the MFT and the SD approaches give
identical results - in our case $\gamma=1/2$. This has led to widespread belief that MFT
and SD are equivalent if the basis for the Fierz ordering is appropriately chosen. However,
this is not the case, as can be seen by calculating also the critical coupling where
spontaneous symmetry breaking sets in in the vector channel (in the absence of other order parameters).
There is again a value $\gamma=-(\bar{\lambda}_{\sigma}+\bar{\lambda}_{V})/(2\bar{\lambda}_{V})$
where MFT and SD give identical results, but it differs from $\gamma=1/2$ as encountered in
the scalar channel\footnote{Actually, $\gamma$ is negative and therefore outside the
range of strict validity of MFT.}. We conclude that there is no possible choice of $\gamma$ where
\emph{both} critical couplings for SSB in the scalar and vector channels are identical
in the MFT and SD approaches. The conceptual and practical difference between the two approaches
appears even more clearly if we consider a model with
eight-fermion-couplings instead of a quartic coupling. Whereas in MFT the onset of SSB can be computed
in one-loop order, only a three loop diagram contributes to the gap equation for $M_{\textrm{F}}$ in
the SD approach. We also warn that the choice of $\gamma$ for which MFT and SD coincide
in a given channel is not necessarily the optimal choice. Comparing the SD result
(or the MFT result for $\gamma=1/2$), namely $\bar{\lambda}_{\sigma}=19.48$, with the perhaps more precise
result from the renormalization group, $\bar{\lambda}_{\sigma}=14.62$, we see that an
''optimal choice'' of the FT for the MFT approach may correspond to a value of $\gamma$ above $1/2$.

\begin{table}[!t]
\begin{center}
\scalebox{0.94}[0.94]{
\begin{tabular}{|c|c|c|c|c|c|c|c}
  \hline
  Approximation & Sect.  & $\gamma=0$ & 0.25 & 0.5 & 0.75 & 1 \\
  \hline
  MFT &\ref{sec::mean} & 39.48 & 38.48 & 37.48 & 36.48 & 35.48 \\
  Ferm. RG & \ref{sec::fermion}  & 41.54 & 41.54 & 41.54 & 41.54 & 41.54 \\
  Bos. RG & \ref{sec::bosoflow}& 36.83 & 36.88 & 36.95 & 37.02 & 37.12 \\
  Adapted Bos. RG &\ref{sec::redef} & 41.54 & 41.54 & 41.54 & 41.54 & 41.54 \\
  \hline
  SD &\ref{sec::schwinger} &   37.48 & 37.48 & 37.48 & 37.48 & 37.48 \\
  \hline
\end{tabular}}
\end{center}
\caption{Critical values $\bar{\lambda}^{\textrm{crit}}_{\sigma}$ for $\bar{\lambda}_{V}=2$ and for
various values of the unphysical
\mbox{parameter $\gamma$} (with $\Lambda=1$).
Progressing from MFT to the bosonic RG and adapted bosonic RG the dependence on $\gamma$ decreases as more
and more diagrams are included.
The
Schwinger-Dyson result is independent of $\gamma$ but contains no vertex corrections in contrast to the RG-calculations.}
\label{tab::crit}
\end{table}
\begin{table}[!t]
\begin{center}
\scalebox{0.94}[0.94]{
\begin{tabular}{|c|c|c|c|c|c|c|}
  \hline
  Approximation & Sect. &  $\gamma=0$ & 0.25 & 0.5 & 0.75 & 1 \\
  \hline
  MFT &\ref{sec::mean}& 39.48 & 29.48 & 19.48 & 9.48 & -0.52 \\
  Ferm. RG &\ref{sec::fermion} & 14.62 & 14.62 & 14.62 & 14.62 & 14.62 \\
  Bos. RG &\ref{sec::bosoflow} & 15.44 & 13.39 & 13.45 & 15.55 & 19.46 \\
  Adapted Bos. RG &\ref{sec::redef}& 14.62 & 14.62 & 14.62 & 14.62 & 14.62 \\
  \hline
  SD &\ref{sec::schwinger}&   19.48 & 19.48 & 19.48 & 19.48 & 19.48 \\
  \hline
\end{tabular}}
\end{center}
\caption{The same\protect\footnote{The negative sign for the critical coupling at $\gamma=1$ in the MFT calculation means that
the system is in the broken phase for any positive value of $\bar{\lambda}_{\sigma}$ in this calculation.}
as in Tab. \ref{tab::crit} but with $\bar{\lambda}_{V}=20$.}
\label{tab::crit2}
\end{table}

\begin{table}[!t]
\begin{center}
\scalebox{0.94}[0.94]{\begin{tabular}{|c|c|c|c|c|c|c|c}
  \hline
  Approximation & Chap.  & $\gamma=0.1$ & 0.25 & 0.5 & 0.75 & 0.9 \\
  \hline
  MFT &\ref{sec::mean} & 74.96 & 68.96 & 58.96 & 48.96 & 42.96 \\
  SD &\ref{sec::schwinger} &   58.96 & 58.96 & 58.96 & 58.96 & 58.96 \\
  Bos. RG=(1) & \ref{sec::bosoflow}& 53.16 & 52.93 & 53.32 & 54.64 & 55.88 \\
  Ad. Bos. RG=(1)+(2)& \ref{sec::fermion}  & 58.83 & 58.83 & 58.83 & 58.83 & 58.83 \\
\hline
  (1)+(3) & \ref{sec::extension} & 53.90 &  53.66& 54.00 & 55.23  & 56.37 \\
  (1)--(3)  & \ref{sec::extension} & 58.14 & 58.04 & 57.88 & 57.73 & 57.64 \\
  (1)--(4)  & \ref{sec::extension} & 61.60 & 61.69 & 61.82 & 61.91 & 61.94 \\
\hline
\end{tabular}}
\end{center}
\caption{Critical coupling $\bar{\lambda}^{\textrm{crit}}_{\sigma}$ for $\bar{\lambda}_{V}=20$
for a UV regularization by the ERGE scheme
We show different approximation steps:
(1) the pointlike contibutions to the mass and the Yukawa coupling (Figs. \ref{fig::mass}, \ref{fig::vertex}),
(2) the pointlike contributions from the box diagrams (Fig. \ref{fig::box}),
(3) the contribution to the WFR from the purely bosonic diagram (Fig. \ref{fig::mass}) and
(4) the contribution to the WFR from the momentum dependence of the diagrams \ref{fig::vertex} and \ref{fig::box}
(those contribute again via an appropriate adaption of the flow).}
\label{tab::crit4}
\end{table}

Partial bosonization is a very powerful tool for understanding strongly interacting fermionic systems
beyond the level of MFT or SD-equations. It allows us to treat the bosonic fluctuations in an explicit
manner and provides for a rather simple framework for the discussion of SSB. Most importantly,
it permits the direct exploration of the ordered phase which is, in practice, almost
inaccessible for the fermionic RG. In order to permit a simple comparison with the fermionic
RG we have used a very crude approximation for the purely bosonic sector by retaining only
a mass term and neglecting bosonic interactions as well as the momentum dependence
of the bosonic propagator. In this approximation the effect of the boson exchange between fermions
does not go beyond pointlike fermionic interactions. Taking into
account only the running of the Yukawa couplings (Fig. \ref{fig::vertex}) in the bosonic RG
of sect. \ref{sec::bosoflow}, we observe already a very substantial decrease of the
Fierz ambiguity as compared to MFT.
The dependence on $\gamma$ is greatly reduced and the numerical value of the critical coupling
comes already close to the result of the fermionic RG. These features can be compared to the inclusion
of higher loop effects in perturbation theory in particle physics: they often reduce the dependence
of the results on unphysical parameters, such as the choice of the renormalization scale.

As compared to perturbation theory, the box diagrams (\ref{fig::box}) are still missing in
the discussion of sect. \ref{sec::bosoflow}. This shortcoming is cured by the
adapted bosonic renormalization group discussed in sect. \ref{sec::redef}. Here
the relation between the bosonic composite fields and the fermion bilinears becomes scale
dependent. This formulation is well adapted to the basic idea of renormalization where
only effective degrees of freedom at a certain scale $k$ and their effective couplings
should matter for physics associated with momenta $q^2\lesssim k^2$. The system should loose all
memory of the detailed microscopic physics. In particular, the choice of an
optimal bosonic field for the long distance physics should not involve the parameters of the
microscopic theory, but rather the renormalized parameters at the scale $k$. In this
formulation it has also become apparent that the distinction between
''fundamental degrees of freedom'' and ''bound states'' becomes a matter of scale \cite{Gies:2002nw}.
The adapted bosonic RG reproduces in our crude approximation the results of the fermionic RG.
We argue that for precision estimates in the partially bosonized approach the
''adaptation'' of the definition of the composite field seems mandatory.

It seems plausible to us that within the local interaction approximation considered in this
paper the most reliable results are obtained
from the fermionic or adapted bosonic RG of \mbox{sects. \ref{sec::fermion} and \ref{sec::redef}.}
First, these methods sum over a larger class of diagrams. The diagrams included by the other approaches
are all contained in the ones taken into account by the fermionic or adapted bosonic RG. Second, the
renormalization procedure accounts properly for the fact that the relevant physics depends on scale.
In the low momentum region relevant for spontaneous symmetry breaking all physics should be
describable in terms of effective low energy couplings. This is not realized by the MFT or SD
approaches where the microscopic or ''bare'' couplings appear explicitly.
We point out, however, that beyond the local interaction approximation the relative
merits of the various methods discussed here depend on the physical situation of the investigated model.

The aim of the investigation of the bosonic RG in
this paper is, of course, not simply a reproduction of the results of the fermionic RG.
By a careful comparison between different approaches we rather want to open the door for
future more elaborate techniques for the study of strongly interacting fermion systems.
With the present results the adapted non-perturbative flow equations in the partially bosonized
approach offer an ideal starting point for an investigation
of spontaneous symmetry breaking. Without too much effort we can now include the
momentum dependence of the bosonic propagator which goes beyond the approximation of
local fermionic interactions. This is crucial for the understanding of critical
phenomena for which the renormalized boson mass vanishes.
We have made a first step in this direction by including the
wave function renormalization in sect. \ref{sec::extension}.
Furthermore, bosonic interactions
can now be included in the form of an effective potential for the scalar field. This is
mandatory for the RG-investigation of the ordered phase where the minimum of the
potential occurs for $\phi\neq 0$. Both effects have already been treated for the NJL-model
with three colors and two flavors for the fermions (quarks) \cite{Berges:1999eu,Jungnickel:1995fp}.
In the view of the results of the present paper, the quantitative accuracy and conceptual setting
of such investigations could further be improved by the ''adaptation'' of the effective
composite variables. Including these effects one by one will also permit a more
detailed appreciation of the uncertainties remaining in the present truncation of the flow.

In summary, some of the methods proposed to deal with
strongly interacting fermionic systems have an additional source of systematic uncertainty.
The Fierz ambiguity is related to the choice of the bosonic or mean field, parametrized by an
unphysical parameter $\gamma$. For those methods the spread of the results within an acceptable
range of $\gamma$ should be considered as a lower bound for this additional
systematic uncertainty. Different values of the parameter $\gamma$ correspond
to an \emph{identical} initial fermionic action. Therefore, vanishing or at least smallness
of the Fierz ambiguity should be required for the self-consistency of an approximation.
We find that, depending on the model and parameters,
MFT can have a very substantial ambiguity which should then be reduced by systematic
improvements.

On the other hand, the Fierz ambiguity
is, of course, not the only source of error - several methods as SD or the fermionic RG
have no such ambiguity by construction. Without a systematic error analysis which is
highly difficult for non-perturbative systems and beyond the scope of this paper
there is no simple overall criterion in order to judge which method is most suitable
- the answer may actually depend on the detailed model and problem. The
non-perturbative flow equations based on an exact renormalization group equation
offer interesting prospects for an understanding of problems where a large
correlation length is important. It is reassuring that the analysis of the
potential Fierz ambiguity in a partially bosonized setting shows that
this approach can be considered as a promising extension beyond MFT.

\begin{acknowledgements}
The authors thank J. Berges, M. Doran, H. Gies, \mbox{F. H\"ofling}, and C. Nowak for
valuable discussions.
\end{acknowledgements}

\section*{Note added}
A renormalization group approach close to the spirit of the SD-equations has
been proposed recently \cite{Wetterich:2002ky} in the context of the
''bosonic effective action''. There the long range bosonic fluctuations can
be incorporated without invoking partial bosonization. The Fierz ambiguity
is therefore absent in this approach.

\begin{appendix}
\section{Threshold Functions for Finite UV-Cutoff}\label{app::cutoff}
In this appendix we compute the threshold functions as defined in \cite{Berges:2000ew}
for the linear cutoff given in \cite{Litim:2000ci}
in the presence of
a finite UV-cutoff $\Lambda$.
The inverse (massless) average propagator $P_{B}$ for bosons and the corresponding
squared quantity $P_{F}$ for fermions are given by
\begin{eqnarray}
P_{B}&=&q^2+Z^{-1}_{\phi,k}R_{k}(q)=q^2(1+r_{B}(q^2)),\quad
\\\nonumber
P_{F}&=&q^2(1+r_{F}(q^2))^2,
\end{eqnarray}
where $r_{B}$ and $r_{F}$ reflect the presence of the IR-cutoff.
The dimensionless functions $r_{B}$ and $r_{F}$ only depend on $y=q^2/k^2$.
For the linear cutoff \cite{Litim:2000ci} they read
\begin{eqnarray}
\label{equ::cutoff}
r_{B}(q,k)&=&\left(\frac{1}{y}-1\right)\Theta(1-y),
\\\nonumber
r_{F}(q,k)&=&\left(\frac{1}{\sqrt{y}}-1\right)\Theta(1-y).
\end{eqnarray}

In presence of an ultraviolet cutoff $\Lambda$ and in the absence of mass terms the threshold functions
can only depend on the ratio $s=k^{2}/\Lambda^{2}$. With
\begin{equation}
\tilde{\partial}_{t}=\frac{q^2}{Z_{\phi,k}}\frac{\partial[Z_{\phi,k}r_{B}]}{\partial t}\frac{\partial}{\partial P_{B}}
+\frac{2}{Z_{\psi,k}}\frac{P_{F}}{1+r_{F}}
\frac{\partial[Z_{\psi,k}r_{F}]}{\partial t}\frac{\partial}{\partial P_{F}}
\end{equation}
we find for bosons ($x=q^2$),
\begin{eqnarray}
&l^{(B)d}_{0}&(\omega,\eta_{\phi},s)
\\\nonumber
&&=\frac{1}{2}k^{-d}\int^{\Lambda^{2}}_{0}dx\,x^{\frac{d}{2}-1}\tilde{\partial_{t}}\ln(P_{B}(x)+\omega k^{2})
\\\nonumber
&&=\frac{2}{d}\left[1-\frac{\eta_{\phi}}{d+2}\right]\frac{1}{1+\omega}\Theta(1-s)
\\\nonumber
&&+\frac{2}{d}s^{-\frac{d}{2}}\left[1-\frac{\left(2+d(1-s^{-1})\right)\eta_{\phi}}{2(d+2)}\right]\frac{1}{1+\omega}
\Theta(s-1)
\end{eqnarray}
and for fermions,
\begin{eqnarray}
&l^{(F)d}_{0}&(\omega,\eta_{\psi},s)
\\\nonumber
&&=\frac{1}{2}k^{-d}\int^{\Lambda^{2}}_{0}dx\,x^{\frac{d}{2}-1}\tilde{\partial_{t}}\ln(P_{F}(x)+\omega k^{2})
\\\nonumber
&&=\frac{2}{d}\left[1-\frac{\eta_{\psi}}{d+1}\right]\frac{1}{1+\omega}\Theta(1-s)
\\\nonumber
&&+\frac{2}{d}s^{-\frac{d}{2}}\left[1-\frac{(d+1-ds^{-1})\eta_{\psi}}{d+1}\right]\frac{1}{1+\omega}\Theta(s-1).
\end{eqnarray}
Higher threshold functions can be obtained simply by differentiating with respect to $\omega$:
\begin{equation}
l^{d}_{n+1}(\omega,\eta,s)=-\frac{1}{n+\delta_{n,0}}\frac{d}{d\omega}l^{d}_{n}(\omega,\eta,s).
\end{equation}
For a finite value of the UV-cutoff $\Lambda$ the threshold functions are
explicitly $s$- and therefore $k$-dependent.
Taking $\Lambda\rightarrow\infty$ we have $s=0$ for any value of $k$. This renders the threshold functions
$k$-independent.
In the present work we neglect the anomalous dimensions $\eta_{\phi}$, $\eta_{\psi}$ and effectively only
consider a fermionic cutoff since $Z_{\phi,k}=0$. For our purpose the fermions are massless and we abbreviate
for $\omega=0$,
\begin{equation}
l^{d}_{n}(0,0,s)=l^{d}_{n}(s).
\end{equation}
This yields explicitly
\begin{equation}
l^{(F),4}_{1}(s)=\frac{1}{2}\left[\Theta(1-s)+s^{-2}\Theta(s-1)\right].
\end{equation}
To obtain the perturbative result from the fermionic RG-equation we used
\begin{equation}
\label{equ::universalint}
\int^{\infty}_{-\infty}dt
k^{2}l^{(F),4}_{1}(s)=\int^{\infty}_{0}dk\, k
l^{(F),4}_{1}(s)=\frac{\Lambda^{2}}{2}.
\end{equation}
As long as we keep the sharp momentum cutoff at $q=\Lambda$ this
integral is universal, i.e. it does not depend on the precise
choice of the IR-cutoff. Indeed the universality is necessary to
reproduce perturbation theory for every choice of the IR-cutoff.

We have also used other cutoff functions $R_{k}$ different from
the linear cutoff. Within the local interaction approximation
we have found that the value of the critical coupling comes out independent
of the choice of $R_{k}$. The basic reason is that a multiplicative change of $l^{(F),4}_{1}$
due to the use of another threshold function can be
compensated by a rescaling of $k$ (cf. Eq. \eqref{equ::fermionflow}). The
rescaling is simply multiplicative for $s<1$, with a suitable generalization for
$s>1$. Critical values of the flow which are defined for $k\to\infty$ are not affected
by the rescaling. Let us demonstrate this for $\bar{\lambda}_{V}$.
Writing Eq. \eqref{equ::fermionflow} in the scale variable $k$ we have
\begin{eqnarray}
\partial_{k}\bar{\lambda}_{V,k}
=-4v_{4}l^{(F),4}_{1}(s)k(\bar{\lambda}_{\sigma,k}+\bar{\lambda}_{V,k})^{2}.
\end{eqnarray}
Rescaling to
\begin{equation}
\tilde{k}(k)=\int^{k}_{0} dk\, k l^{(F),4}_{1}(s)
\end{equation}
we find
\begin{equation}
\partial_{\tilde{k}}\bar{\lambda}_{V,\tilde{k}}
=-4v_{4} (\bar{\lambda}_{\sigma,\tilde{k}}+\bar{\lambda}_{V,\tilde{k}})^{2}.
\end{equation}
Due to the universality of Eq. \eqref{equ::universalint} the
domain for $k$, $[0,\infty]$, is now mapped to
$[0,\frac{\Lambda^{2}}{2}]$, giving the domain for $\tilde{k}$
independent of the IR-cutoff. Having obtained identical
differential equations for every choice of the IR-cutoff without any
rescaling of $\bar{\lambda}$ establishes the above claim for the
critical couplings.

Note however, that this would not hold if we would start the integration
of the flow equation at
$k=\Lambda$. In this case the domain $[0,\Lambda]$ for $k$ is mapped
into an interval for $\tilde{k}$ that depends on the threshold function
and therefore on $R_{k}$. Actually,
the $R_{k}$ dependence in this case is not very surprising
because different IR-cutoffs then correspond to different UV regularizations (see App. \ref{app::ergescheme}).
Since our model is naively non-renormalizable results can depend
on the choice of UV regularization.

\section{UV Regularization -- The ERGE Scheme}\label{app::ergescheme}
\begin{figure}[!t]
\begin{center}
\begin{picture}(220,150)(10,10)
\Text(210,35)[]{\scalebox{1.3}[1.3]{$\frac{q^2}{\Lambda^{2}}$}}
\includegraphics[width=8cm]{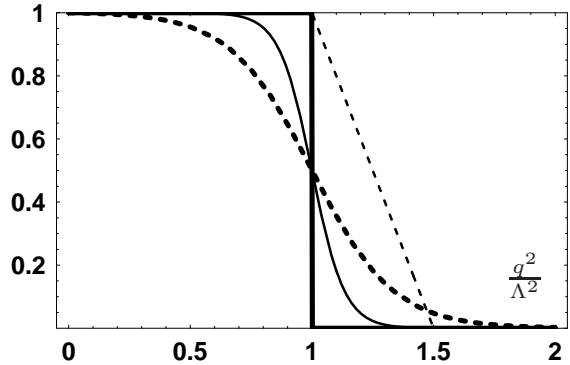}
\end{picture}
\end{center}
\caption{In a UV regularized theory not all modes contribute completely.
This plot schematically depicts ``how much'' each mode contributes.
The thick line is for the sharp momentum cutoff. All modes with $q^{2}\leq\Lambda^{2}$ are included completely.
Other UV regularizations (dashed, thin dashed and thin solid line) typically not only include a small fraction
of the high momentum modes, $q^2>\Lambda^{2}$, but in addition leave out a small fraction
of the low momentum modes.}
\label{fig::regu}
\end{figure}

In the first eight sections we have implemented the UV regularization by a sharp cutoff in all integrals
over momentum space. This
is often not the most practical regularization.
As an alternative, the modes with $q^2>\Lambda^2$ are not completely left out, only suppressed, as
depicted in Fig. \ref{fig::regu}.

This can be realized by a modification of the propagator in the action
yielding an ``UV regularized classical action''. Different UV regularizations usually correspond
to different ``classical actions''. Therefore, it is no surprise, that different
UV regularizations give different results. In particular, this
is true for the critical coupling in the NJL model, Eq. \eqref{equ::faction}, as one
can see by comparing Tabs. \ref{tab::crit}, \ref{tab::crit2}
calculated using a sharp UV cutoff at $\Lambda$ with Tab. \ref{tab::crit4},
which employs a UV regularization by the ERGE scheme (below) with the linear cutoff Eq. \eqref{equ::cutoff}.
One expects that dimensionless low energy quantities such as the ratio of particle masses and
order parameters in the phase with spontaneous symmetry breaking, show much less dependence
on the regularization \cite{Berges:1999eu}.

In App. \ref{app::cutoff} we have evaluated the threshold functions for a theory
which is UV regularized by a sharp cutoff in momentum space. The threshold
functions depend on the ratio $s=\frac{k^2}{\Lambda^2}$ in a rather complicated way.
Cutoff scale independent threshold functions would be desirable, to simplify
numerical calculations.

This is implemented by the ERGE regularization \cite{Ellwanger:1997tp,Bergerhoff:1997cv} where $\Lambda$ is taken to
infinity in the definition of the threshold functions (i.e. $s=0$). On the other hand,
the classical action is now specified indirectly by the ``initial value'' of the effective average
action $\Gamma_{\Lambda}$ at some ultraviolet scale $\Lambda$. Depending on the choice
of the cutoff function $R_{k}(q)$ the fluctuations with momenta $q^2>\Lambda^2$ have
not yet been integrated out completely, as depicted in Fig. \ref{fig::regu}. As a result,
the relation between ``classical couplings'' $\bar{\lambda}_{\sigma}$, $\bar{\lambda}_{V}$ and
physical observables depends on the choice of $R_{k}$ and differs from the regularization with
a sharp cutoff. We emphasize that this is a difference in the \emph{definition} of the
classical couplings and should not be confounded with approximations errors.

Although it is usually not the simplest method, we can invoke UV regularization by the
ERGE scheme also in the context of perturbation theory or SDE. This follows
along the lines indicated in Sect. \ref{sec::fermion} for perturbation theory.
Typically any expression can be written in terms of inverse propagators $P$, internal
momenta $q$ we integrate over, and external momenta $p$ we do not integrate over,
\begin{equation}
\label{equ::unreg}
\int_{q} F(P,q,p).
\end{equation}
A specific ERGE scheme is specified by the choice of the IR regulator $R_{k}$.
Replacing the inverse propagator $P$ by the IR regularized inverse propagator \mbox{$P+R_{k}$}
we can calculate the contribution from each \mbox{scale $k$,} \mbox{$k^{-1}\tilde{\partial}_{t}F(P+R_{k},q,p)$.}
Integrating over all scales from \mbox{$k_{0}=\Lambda$} to $k=0$ we obtain the UV regularized
expression,
\begin{equation}
\int^{0}_{k_{0}=\Lambda} dk\,k^{-1}\tilde{\partial_{t}}\left[\int_{q} F(P+R_{k},q,p)\right].
\end{equation}
\end{appendix}

\end{document}